\documentclass[preprintnumbers, floatfix, letterpaper, onecolumn,aps,prd,epsfig,nofootinbib,natbib,longbibliography]{revtex4-2}
\usepackage{bm,graphicx,dcolumn,epstopdf,epsf, latexsym,mathbbol, amssymb,amsmath,color,slashed, mathrsfs,mathcomp,simplewick}
\usepackage[center]{subfigure}
\usepackage{multirow}
\usepackage{makecell}
\usepackage{placeins}
\usepackage{epstopdf}
\usepackage[colorlinks,linkcolor=blue,citecolor=blue,urlcolor=blue]{hyperref}

\usepackage{amsmath}
\begin{document}
\allowdisplaybreaks
 \newcommand{\bq}{\begin{equation}}
 \newcommand{\eq}{\end{equation}}
 \newcommand{\bqn}{\begin{eqnarray}}
 \newcommand{\eqn}{\end{eqnarray}}
 \newcommand{\nb}{\nonumber}
 \newcommand{\lb}{\label}
\newcommand{\f}{\frac}
\newcommand{\p}{\partial}
\newcommand{\PRL}{Phys. Rev. Lett.}
\newcommand{\PLB}{Phys. Lett. B}
\newcommand{\PRD}{Phys. Rev. D}
\newcommand{\CQG}{Class. Quantum Grav.}
\newcommand{\JCAP}{J. Cosmol. Astropart. Phys.}
\newcommand{\JHEP}{J. High. Energy. Phys.}
\newcommand{\orcid}[1]{\href{https://orcid.org/#1}{\includegraphics[width=10pt]{orcid}}}
\title {Dynamics of Reheating in Modified Loop Quantum Cosmologies: mLQC-I }

\author{Jared Fier}
\email{Jared$\_$Fier@baylor.edu}

\author{Rui Pan}
\email{rui$\_$pan1@baylor.edu}

\author{Jamal Saeed}
\email{Jamal$\_$Saeed1@baylor.edu}

\author{Anzhong Wang}
\email{Anzhong$_$Wang@baylor.edu; the corresponding author}

\affiliation{GCAP-CASPER, Department of Physics $\&$ Astronomy, Baylor University, Waco, TX 76798-7316, USA}
\date{\today}

\begin{abstract}

We study the reheating epoch and the total bounce-to-present expansion
in modified loop quantum cosmology I (mLQC-I), using current
observational determinations of the scalar amplitude $A_s$ and spectral index
$n_s$ from Planck 2018, ACT 2025, and the combined Planck+ACT+DESI DR2 data.
We first consider the power-law plateau (PLP) potential, whose
normalization is determined independently from each observational pair
$(A_s,n_s)$. We find that, for the kinetic-energy-dominated initial
conditions considered here, the quantum-bounce regime is remarkably
insensitive to this normalization, while the subsequent inflationary
and reheating evolution shows a more appreciable dependence. We then
investigate the sensitivity of the results to the inflationary potential
by considering the Starobinsky potential and a polynomial chaotic
potential, the latter admitting two distinct mass branches for the same
$(A_s,n_s)$. The reheating e-fold number and reheating temperature are
found to depend strongly on both the form of the potential and, in the
polynomial case, the branch selected. In contrast, the total bounce-to-present 
expansion satisfies the conservative bound $N_T\gtrsim129$ over the ranges 
considered here and remains close to the characteristic scale $N_T\sim130$ 
found in LQC. Moreover, the difference between mLQC-I and LQC is remarkably stable,
$N_T^{\rm mLQC-I}-N_T^{\rm LQC}\simeq -0.24$, for all three potentials and 
both polynomial branches. These results
suggest that reheating is strongly model dependent, whereas the small
systematic shift in the total expansion is primarily associated with
the different background quantum geometries of LQC and mLQC-I.
\end{abstract}

\maketitle

\section{Introduction}
\label{introduction}

Inflation provides a remarkably successful framework for describing the
very early universe.  In addition to addressing the horizon and flatness
problems of the standard hot Big-Bang model, it provides a mechanism for
generating primordial perturbations whose statistical properties can be
confronted directly with observations of the cosmic microwave background
(CMB) and large-scale structure
\cite{Guth:1980zm,Starobinsky:1980te,Linde:1981mu,Guth:1982ec,
Linde:1983gd,Baumann:2009ds,Planck:2018jri,BICEP:2021xfz,
AtacamaCosmologyTelescope:2025blo,AtacamaCosmologyTelescope:2025nti}.  
However, inflation by itself does not complete
the thermal history of the early universe.  After the accelerated phase
ends, the energy stored in the inflaton must be transferred into
ordinary degrees of freedom, leading eventually to a radiation-dominated
hot universe.  This reheating epoch provides the bridge between
inflation and the standard thermal history and affects the relation
between the inflationary observables and the number of e-folds between
horizon crossing and the end of inflation
\cite{Allahverdi:2010xz,Amin:2014eta,Bassett:2005xm,Lozanov:2019jxc,
Cook:2015vqa,Martin:2014nya}.  Consequently, observational constraints
on $(A_s,n_s)$ can also be used to constrain the duration and
temperature of reheating.

Despite its phenomenological success, the conventional inflationary
description does not resolve the initial singularity and, when extended
far enough into the past, eventually reaches a regime in which quantum
gravitational effects can no longer be neglected.  Relatedly, sufficiently
long inflationary histories can lead to the trans-Planckian problem,
because modes observed today may originate with wavelengths below the
Planck scale when evolved backward in a purely classical spacetime
\cite{Martin:2000xs,Brandenberger:2012aj,Martin:2013tda}.  These issues
motivate a framework in which the Planck regime and the transition to
the inflationary phase are described within a quantum theory of gravity.

Loop quantum cosmology (LQC) provides such a framework for homogeneous
cosmological spacetimes.  In LQC, quantum geometry effects replace the
classical Big-Bang singularity by a nonsingular quantum bounce, joining
a contracting branch to the expanding universe
\cite{Ashtekar:2011ni,Bojowald:2015iga,Agullo2023,Li:2023dwy}.  The
post-bounce dynamics has been studied extensively, and slow-roll
inflation has been shown to arise generically for broad classes of
initial data
\cite{Ashtekar:2009mm,Ashtekar:2011rm}.  This makes LQC particularly
well suited for connecting Planck-scale physics, inflation, and the
subsequent hot Big-Bang evolution in a single dynamical framework.

Modified loop quantum cosmologies provide closely related but distinct
effective realizations of the quantum geometry
\cite{Yang:2009fp,Dapor:2017rwv,Assanioussi:2018hee,Li:2018opr,Li:2018fco,Li:2019ipm,Li:2021mop,Saeed:2024xhk}. 
Among them, modified loop quantum cosmology I (mLQC-I) has received
considerable attention because its effective Hamiltonian and its
pre- and post-bounce branches differ nontrivially from those of standard
LQC.  The genericity of inflation and the characteristic post-bounce
evolution in mLQC-I have been investigated in detail
\cite{Li:2018opr,Li:2018fco,Li:2019ipm,Li:2021mop,Saeed:2024xhk}.  By the onset of slow-roll inflation,
however, the background quantum corrections are already strongly
suppressed, so the principal imprint of mLQC-I on late-time quantities
enters through the mapping between the Planck regime and the
observationally relevant inflationary scales.

Recently, the reheating phase was incorporated systematically into the
LQC framework, showing that the duration and temperature of reheating
can be related to inflationary observables and that the total expansion
from the quantum bounce to the present can be constrained once the
reheating history is taken into account \cite{Yogesh:2024iip}.  This
naturally raises two questions.  First, how are these conclusions
modified when the underlying quantum geometry is changed from LQC to
mLQC-I?  Second, to what extent are the resulting reheating constraints
and the bounce-to-present expansion sensitive to the choice of
inflationary potential?

These are the questions addressed in the present work.  We use the
effective post-bounce dynamics of mLQC-I and normalize the inflationary
potential directly from the observed scalar amplitude and spectral index
at the pivot scale.  We consider the central values inferred from
Planck 2018, ACT 2025, and the combined Planck+ACT+DESI DR2 analysis,
following Refs.~\cite{Brown:2025hcb,Fier:2026xko}.  Since an explicit expression for
the mLQC-I correction factor $F_{\rm I}(k)$ associated with the power spectrum
\cite{Li:2021mop,Pan:2025ppa} is not yet available, we work throughout
at leading order and adopt $F_{\rm I}(k_*)=1$.  This approximation
isolates the effects of the mLQC-I background evolution and provides a
useful leading-order starting point for studying reheating and the total
expansion history.

We first take the power-law plateau (PLP) potential as our reference
model.  Its normalization is determined independently for each observed
pair $(A_s,n_s)$, which allows us to test directly whether the
post-bounce dynamics is sensitive to the observationally inferred
potential scale.  We then compute the reheating e-fold number
$N_{\rm re}$ and reheating temperature $T_{\rm re}$ for representative
constant values of the effective reheating equation of state.  The
total number of e-folds from the bounce to the present,
$N_T=\ln(a_0/a_B)$, is obtained by evolving the observationally fixed
horizon-crossing data backward to the mLQC-I bounce.

Several results emerge from this analysis.  The genuinely quantum
post-bounce regime is only weakly affected by the observational
normalization of the PLP potential, while the later inflationary and
reheating evolution is more sensitive to it.  For the PLP model, 
the total bounce-to-present expansion satisfies
$N_T\gtrsim129$ over the displayed interval and remains close to the
characteristic scale $N_T\sim130$.  Relative to LQC, mLQC-I produces a small systematic
shift,
\[
\Delta N_T
\equiv
N_T^{\rm mLQC-I}-N_T^{\rm LQC}
\simeq -0.24.
\]
To determine whether these conclusions depend strongly on the inflaton
potential, we repeat the analysis for the Starobinsky potential and for
a polynomial chaotic potential \cite{Kallosh:2025ijd}.  The reheating quantities are found to
be quite sensitive to the potential.  In the polynomial model, two
distinct mass solutions associated with the same observational pair
$(A_s,n_s)$ lead to markedly different reheating histories and to a
shift of order one e-fold in the absolute value of $N_T$.  Nevertheless,
for all three potentials and for both polynomial branches, the
bounce-to-present expansion remains of order $N_T\sim130$, while the
LQC--mLQC-I difference stays remarkably close to
$\Delta N_T\simeq-0.24$.  This suggests that the latter shift is tied
primarily to the difference between the two background quantum
geometries rather than to the detailed form of the inflaton potential.

The remainder of the paper is organized as follows.
Section~\ref{sec:background_mLQCI} reviews the effective post-bounce
dynamics of mLQC-I, determines the PLP normalization from current
observations, and studies the dependence of the background evolution on
that normalization.  In Sec.~\ref{sec:reheating_mLQCI} we analyze the
reheating epoch and calculate $N_{\rm re}$ and $T_{\rm re}$.
Section~\ref{sec:total_expansion_mLQCI} determines the total
bounce-to-present expansion and compares the result directly with LQC.
In Sec.~\ref{sec:potential_dependence} we examine the sensitivity of
our conclusions to the Starobinsky and polynomial chaotic potentials.
Our conclusions and remarks are presented in
Sec.~\ref{sec:conclusions}.

\section{Background dynamics in mLQC-I and observational normalization}
\label{sec:background_mLQCI}

In this section, we first summarize the effective post-bounce dynamics of
modified loop quantum cosmology I (mLQC-I), and then determine the
normalization of the power-law plateau (PLP) potential from current CMB
observations. Since in the present work the perturbations are initialized
at the quantum bounce, only the expanding branch, $t\geq t_B$, is needed
for the background evolution considered below. Throughout the numerical
analysis we use units $G=m_{\rm Pl}^{-2}=1$, unless stated otherwise.

\subsection{Effective post-bounce dynamics}
\label{subsec:effective_dynamics}

For a spatially flat FLRW spacetime,
    $ds^2=-dt^2+a^2(t)d{\bf x}^2$,
the effective Hamiltonian of mLQC-I is given by
\begin{equation}
{\cal H}_{\rm I}
=
\frac{3v}{8\pi G\lambda^2}
\left[
 \sin^2(\lambda b)
 -\frac{1+\gamma^2}{4\gamma^2}\sin^2(2\lambda b)
\right]
+{\cal H}_{\rm M},
\label{eq:H_mLQCI}
\end{equation}
where $v=a^3$, $\gamma\simeq0.2375$ is the Barbero--Immirzi parameter \cite{Agullo2023},
$\lambda$ is related to the minimum area gap, and ${\cal H}_{\rm M}=v\rho$.
The canonical variables satisfy
\begin{equation}
    \{b,v\}=4\pi G\gamma.
\end{equation}
For a scalar field $\phi$ with potential $V(\phi)$,
\begin{equation}
    \rho=\frac{1}{2}\dot\phi^2+V(\phi),\qquad
    P=\frac{1}{2}\dot\phi^2-V(\phi),
\label{eq:rhoP}
\end{equation}
and the Klein--Gordon equation retains its standard form,
\begin{equation}
    \ddot\phi+3H\dot\phi+V_{,\phi}=0.
\label{eq:KG}
\end{equation}

The Hamiltonian constraint implies
\begin{equation}
 \rho
 =
 \rho_c\,\sin^2(\lambda b)
 \left[1-(1+\gamma^2)\sin^2(\lambda b)\right],
\label{eq:rho_b}
\end{equation}
where
\begin{equation}
    \rho_c=\frac{3}{8\pi G\gamma^2\lambda^2}.
\end{equation}
The maximum energy density in mLQC-I is
\begin{equation}
    \rho_c^{\rm I}
    =
    \frac{\rho_c}{4(1+\gamma^2)},
\label{eq:rhocI}
\end{equation}
at which the quantum bounce occurs \cite{Li:2018fco,Saeed:2024xhk}.

On the expanding branch, the modified Friedmann equation can be written as
\bqn
H^2
&=&
\frac{8\pi G}{3}\rho
\left(1-\frac{\rho}{\rho_c^{\rm I}}\right)
\left[
1+
\frac{\gamma^2}{1+\gamma^2}
\frac{\rho/\rho_c^{\rm I}}
{\left(1+\sqrt{1-\rho/\rho_c^{\rm I}}\right)^2}
\right],
\qquad t\geq t_B .
\label{eq:Friedmann_mLQCI}
\eqn 
For $\rho/\rho_c^{\rm I}\ll1$, Eq.~(\ref{eq:Friedmann_mLQCI})
reduces to the standard Friedmann equation,
$H^2\simeq 8\pi G\rho/3$. In particular, by the onset of slow-roll
inflation the mLQC-I corrections to the background dynamics are already
negligible \cite{Li:2018fco,Saeed:2024xhk}.
For later use, it is convenient to introduce
\begin{equation}
    x\equiv\frac{\rho}{\rho_c^{\rm I}},
    \qquad
    H^2=\frac{8\pi G\rho_c^{\rm I}}{3}{\digamma}_{\rm I}(x),
\end{equation}
where
\begin{equation}
{\digamma}_{\rm I}(x)
=
x(1-x)
\left[
1+
\frac{\gamma^2}{1+\gamma^2}
\frac{x}{(1+\sqrt{1-x})^2}
\right].
\label{eq:FI_background}
\end{equation}
Here ${\digamma}_{\rm I}(x)$ is the background function entering the
modified Friedmann equation.  It should be distinguished from the
plain ${\cal{F}}_{\rm I}(k)$ introduced below, which parametrizes possible
mLQC-I corrections to the primordial perturbation spectra.
Using the conservation equation
$\dot\rho=-3H(\rho+P)=-3H\dot\phi^2$, we obtain
\begin{equation}
    \dot H
    =
    -4\pi G\dot\phi^2
    \frac{d{\digamma}_{\rm I}}{dx},
\label{eq:Hdot_mLQCI}
\end{equation}
and therefore
\begin{equation}
    \epsilon_H\equiv-\frac{\dot H}{H^2}.
\label{eq:epsilonH}
\end{equation}

We define the onset and end of the ordinary inflationary phase by the
successive post-transition crossings
\begin{equation}
    \epsilon_H(t_i)=1,
    \qquad
    \epsilon_H(t_{\rm end})=1,
\end{equation}
with $\epsilon_H<1$ for $t_i<t<t_{\rm end}$. The corresponding number
of inflationary e-folds is
\begin{equation}
    N_{\rm inf}
    =
    \ln\frac{a(t_{\rm end})}{a(t_i)}.
\label{eq:Ninf}
\end{equation}
We emphasize that $N_{\rm inf}$ denotes the total number of e-folds
during the ordinary inflationary phase, measured from its onset $t_i$
to its end $t_{\rm end}$.  It should not be confused with $N_*$,
introduced below, which counts the e-folds only from the horizon
crossing of the observational pivot mode, $t_*$, to $t_{\rm end}$.

As in the LQC analysis \cite{Yogesh:2024iip}, the scale factor at the
bounce can be fixed by the rescaling symmetry,
\begin{equation}
    a_B\equiv a(t_B)=1,
    \qquad
    t_B=0.
\end{equation}
For any specified value of $\phi_B$, the Hamiltonian constraint determines
$\dot\phi_B$ up to a sign,
\begin{equation}
    \dot\phi_B
    =
    \pm
    \sqrt{2\left[\rho_c^{\rm I}-V(\phi_B)\right]}.
\label{eq:phidotB}
\end{equation}
In the following, we focus on kinetic-energy-dominated initial conditions
at the bounce, for which $V(\phi_B)\ll\rho_c^{\rm I}$.

\subsection{PLP potential and determination of $V_0$}
\label{subsec:V0_data}

We first consider the same PLP potential employed in the LQC reheating
analysis \cite{Yogesh:2024iip},
\begin{equation}
    V(\phi)
    =
    V_0 m_{\rm Pl}^4
    \frac{\phi^2}{\phi^2+m^2},
\label{eq:PLP}
\end{equation}
and set $m=m_{\rm Pl}$. In Planck units this becomes
\begin{equation}
    V(\phi)=V_0\frac{\phi^2}{1+\phi^2}.
\label{eq:PLP_planck}
\end{equation}
The dimensionless constant $V_0$ is not treated as a free parameter.
Instead, it is fixed by the observed scalar amplitude $A_s$ and scalar
spectral index $n_s$ at the pivot scale
$k_*/a_0=0.05\,{\rm Mpc}^{-1}$.

At horizon crossing, the background is already deep in the classical
slow-roll regime. In the present leading-order treatment we adopt
\begin{equation}
    {\cal{F}}_{\rm I}(k_*)=1,
    \qquad
    \xi_{\rm I}(k_*)\equiv
    \left.
    \frac{d\ln {\cal{F}}_{\rm I}}{d\ln k}
    \right|_{k_*}
    =0,
\label{eq:FI_perturb_approx}
\end{equation}
where ${\cal{F}}_{\rm I}(k)$ denotes the possible mLQC-I correction to the
primordial spectra.  In contrast to the LQC case, an explicit expression
for ${\cal{F}}_{\rm I}(k)$ for the fourth-order adiabatic state imposed at the
mLQC-I bounce \cite{Li:2021mop,Pan:2025ppa} is not presently available.
We therefore adopt
Eq.~(\ref{eq:FI_perturb_approx}) as a leading-order approximation.  Since
the observational pivot mode exits the Hubble radius when the background
mLQC-I corrections are already extremely small, we expect this
approximation to provide a useful first estimate; nevertheless, its
accuracy should ultimately be checked by an explicit perturbative
calculation of ${\cal{F}}_{\rm I}(k)$.

To leading order in slow roll,
\begin{equation}
    n_s-1
    \simeq
    -2\left(3\epsilon_V-\eta_V\right),
    \qquad
    r_*=16\epsilon_V^*,
\label{eq:nsr_slowroll}
\end{equation}
where, for the PLP potential,
\begin{equation}
    \epsilon_V
    =
    \frac{1}{4\pi(\phi+\phi^3)^2},
    \qquad
    \eta_V
    =
    \frac{1-3\phi^2}
    {4\pi(\phi+\phi^3)^2}.
\label{eq:eps_eta_PLP}
\end{equation}
For a measured value of $n_s$, Eq.~(\ref{eq:nsr_slowroll}) uniquely
determines $\phi_*$ on the branch considered here. The scalar amplitude
then gives
\begin{equation}
    H_*^2
    =
    \pi m_{\rm Pl}^2\,\epsilon_V^* A_s,
\label{eq:Hstar_As}
\end{equation}
while the slow-roll Friedmann equation yields
\begin{equation}
    H_*^2
    \simeq
    \frac{8\pi}{3m_{\rm Pl}^2}V(\phi_*).
\label{eq:Hstar_V}
\end{equation}
Combining Eqs.~(\ref{eq:Hstar_As}) and (\ref{eq:Hstar_V}), we find
\begin{equation}
    V_0
    =
    \frac{3A_s\epsilon_V^*}{8}
    \frac{1+\phi_*^2}{\phi_*^2}.
\label{eq:V0_from_data}
\end{equation}

Following Ref.~\cite{Brown:2025hcb}, we consider the central values from
Planck 2018, ACT 2025, and the combined Planck+ACT+DESI DR2 data set:
\begin{align}
(A_s,n_s)_{\rm P}
&=
(2.0989\times10^{-9},\,0.9649),
\\
(A_s,n_s)_{\rm ACT}
&=
(2.1179\times10^{-9},\,0.9666),
\\
(A_s,n_s)_{\rm C}
&=
(2.1370\times10^{-9},\,0.9752).
\label{eq:data_sets}
\end{align}
The resulting values of $\phi_*$, $r_*$, $H_*$ and $V_0$ are listed in
Table~\ref{tab:V0_sets}. The Planck and ACT determinations are close to each other, whereas the
combined data yield a noticeably smaller normalization of the PLP
potential. It is therefore useful to examine explicitly whether the
post-bounce evolution is sensitive to these differences.

\FloatBarrier
\begin{table}[!ht]
\centering
\caption{PLP parameters inferred from the three observational data sets,
assuming $F_{\rm I}(k_*)=1$. We use $m_{\rm Pl}=1$.}
\label{tab:V0_sets}
\begin{tabular}{lcccc}
\hline\hline
Data set
& $\phi_*$
& $r_*$
& $H_*$
& $V_0$
\\
\hline
Planck 2018
& 1.751892
& 0.025055
& $3.2133\times10^{-6}$
& $1.6341\times10^{-12}$
\\
ACT 2025
& 1.777810
& 0.023272
& $3.1109\times10^{-6}$
& $1.5206\times10^{-12}$
\\
Combined
& 1.939285
& 0.014937
& $2.5035\times10^{-6}$
& $9.4706\times10^{-13}$
\\
\hline\hline
\end{tabular}
\end{table}

\subsection{Dependence of the post-bounce evolution on $V_0$}
\label{subsec:background_V0_dependence}

To isolate the effect of the potential normalization, we first evolve the
same bounce initial condition for all three values of $V_0$.  This common
initial condition is introduced only as a controlled sensitivity test of
the background dynamics; it is not imposed as the self-consistent
horizon-crossing trajectory associated with each observational data set.
As a representative example, we choose
    $\phi_B=-4$,
    $\dot\phi_B>0$,
where $\dot\phi_B$ is fixed by Eq.~(\ref{eq:phidotB}). For
$\gamma=0.2375$ and $\rho_c=0.41\,m_{\rm Pl}^4$,
    $\rho_c^{\rm I}
    \simeq
    9.7027\times10^{-2}\,m_{\rm Pl}^4$.

\begin{figure*}[t]
\centering
\includegraphics[width=0.95\textwidth]{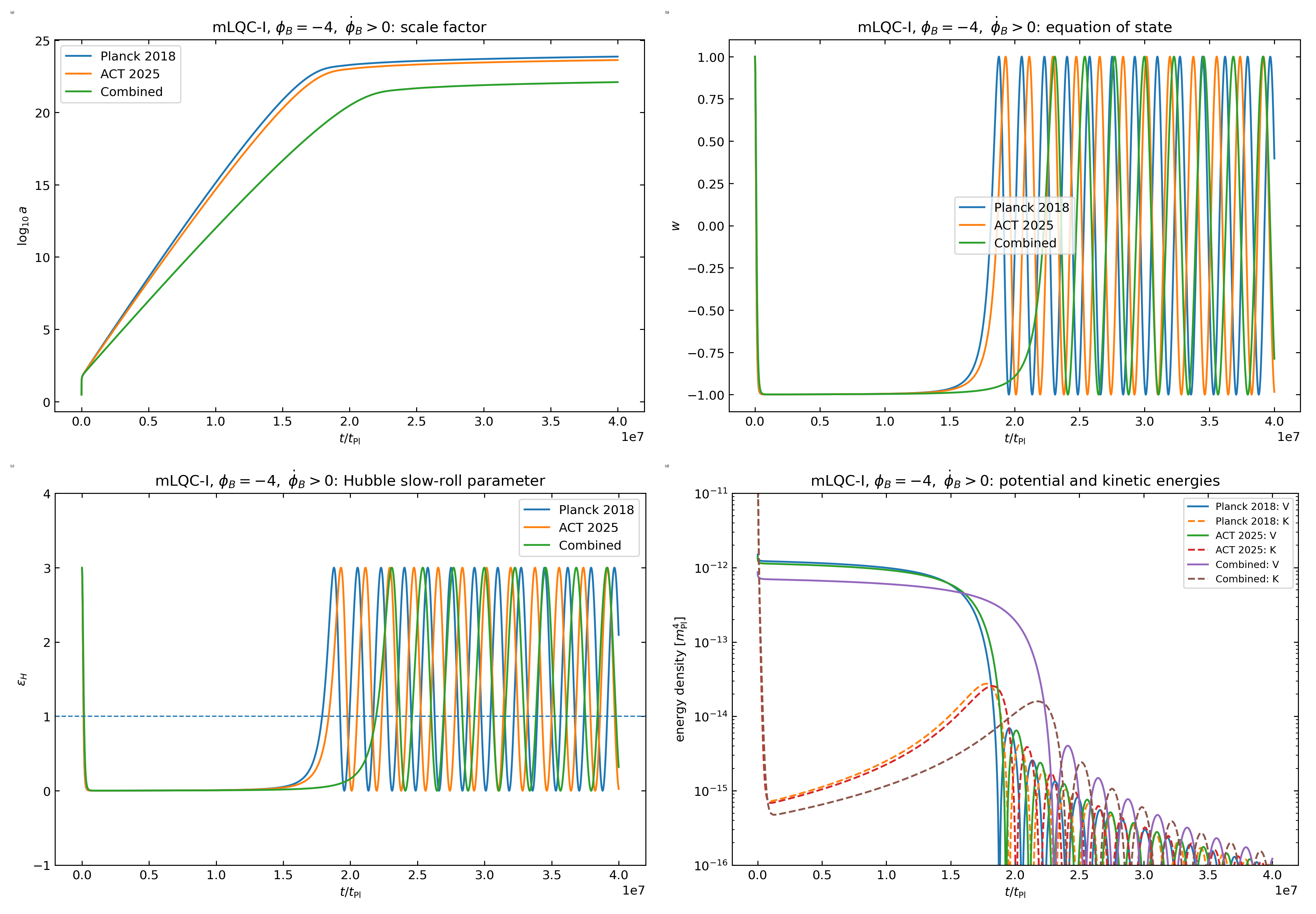}
\caption{Post-bounce background evolution in mLQC-I for
$\phi_B=-4$ and $\dot\phi_B>0$, using the three values of $V_0$
listed in Table~\ref{tab:V0_sets}. The panels show the scale factor,
equation-of-state parameter, Hubble slow-roll parameter, and kinetic and
potential energies. The three solutions essentially coincide near the
bounce and begin to separate only during the transition toward inflation.}
\label{fig:background_threeV0}
\end{figure*}

Figure~\ref{fig:background_threeV0} compares the resulting evolution of
the scale factor, equation of state, Hubble slow-roll parameter, and scalar
kinetic and potential energies. The three solutions are practically
indistinguishable in the immediate vicinity of the quantum bounce. This is
expected because all three values of $V_0$ are of order
$10^{-12}m_{\rm Pl}^4$, whereas the bounce energy density is of order
$10^{-1}m_{\rm Pl}^4$. Consequently, for kinetic-energy-dominated initial
conditions the bounce dynamics is controlled almost entirely by the
quantum geometry.

Quantitatively, for the above chosen initial condition, 
the spread among the three solutions is
\begin{equation}
\left.
\frac{\Delta a}{a}
\right|_{t=10^2t_{\rm Pl}}
\simeq3.5\times10^{-7},
\qquad
\left.
\frac{\Delta\rho}{\rho}
\right|_{t=10^2t_{\rm Pl}}
\simeq3.8\times10^{-9},
\end{equation}
with
\begin{equation}
\left.
\Delta w
\right|_{t=10^2t_{\rm Pl}}
\simeq4.7\times10^{-6}.
\end{equation}
Even at $t=10^4t_{\rm Pl}$ the three solutions remain very close. Their
differences become appreciable only later, when the kinetic energy has
redshifted sufficiently for the inflationary potential to become
dynamically important. The characteristic times and total number of inflationary e-folds for the
same initial condition are  presented in Table \ref{tab:background_times}.

\begin{table}[!ht]
\centering
\caption{Characteristic post-bounce quantities for
$\phi_B=-4$ and $\dot\phi_B>0$. Here $t_{w=0}$ denotes the first
post-bounce time at which the scalar-field equation of state crosses
$w=0$, while $t_i$ and $t_{\rm end}$ denote the onset and end of the
ordinary inflationary phase.}
\label{tab:background_times}
\begin{tabular}{lcccc}
\hline\hline
Data set
& $t_{w=0}/t_{\rm Pl}$
& $t_i/t_{\rm Pl}$
& $t_{\rm end}/t_{\rm Pl}$
& $N_{\rm inf}$
\\
\hline
Planck 2018
& $9.0237\times10^4$
& $1.18190\times10^5$
& $1.78969\times10^7$
& 48.5212
\\
ACT 2025
& $9.3612\times10^4$
& $1.22617\times10^5$
& $1.83834\times10^7$
& 47.9976
\\
Combined
& $1.19210\times10^5$
& $1.56213\times10^5$
& $2.19236\times10^7$
& 44.6606
\\
\hline\hline
\end{tabular}
\end{table}

Thus, the Planck and ACT normalizations lead to very similar post-bounce
histories, whereas the smaller $V_0$ preferred by the combined data delays
the transition toward potential domination and modestly reduces the
inflationary duration for this particular bounce initial condition.
The values of $N_{\rm inf}$ in Table~\ref{tab:background_times} should
therefore not be identified with the observationally inferred $N_*$.
When the horizon-crossing data are instead evolved backward
self-consistently to the bounce, the three central observational data sets
give, on the branch $\dot\phi_B>0$,
\begin{equation}
 \phi_B \simeq -4.0441,\qquad -4.0747,\qquad -4.2682,
\label{eq:selfconsistent_phiB}
\end{equation}
for Planck 2018, ACT 2025, and the combined data, respectively.  By the
symmetry of the PLP potential, these are equivalent to the solutions with
positive $\phi_B$ and $\dot\phi_B<0$.

\subsection{Inflationary e-folds as a function of the bounce data}
\label{subsec:Ninf_phiB}

\begin{figure*}[t]
\centering
\includegraphics[width=0.95\textwidth]{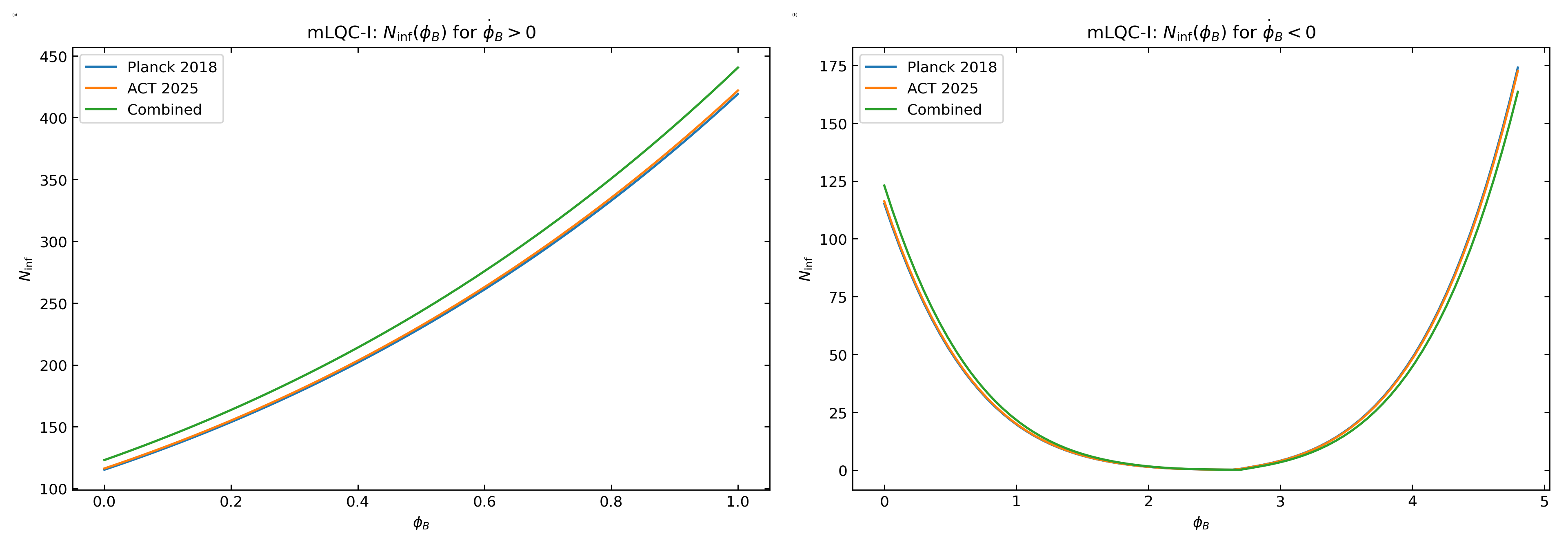}
\caption{Number of inflationary e-folds $N_{\rm inf}$ as a function of
$\phi_B$ for the three observationally determined values of $V_0$.
Panel (a) corresponds to $\dot\phi_B>0$, while panel (b) corresponds to
$\dot\phi_B<0$. Planck and ACT lead to very similar predictions, whereas
the combined data give the largest deviation.}
\label{fig:Ninf_threeV0}
\end{figure*}

To further quantify the dependence on the potential normalization, we
calculate $N_{\rm inf}$ as a function of $\phi_B$ for both signs of
$\dot\phi_B$. The results are shown in
Fig.~\ref{fig:Ninf_threeV0}. For $\dot\phi_B>0$, the total number of
inflationary e-folds grows rapidly with $\phi_B$. For example,
\begin{align}
\phi_B=0:
&\qquad
N_{\rm inf}
=
115.21,\;116.22,\;123.03,
\\
\phi_B=0.5:
&\qquad
N_{\rm inf}
=
229.97,\;231.71,\;243.39,
\\
\phi_B=1:
&\qquad
N_{\rm inf}
=
419.20,\;421.97,\;440.53,
\end{align}
for Planck, ACT, and the combined data, respectively.

For $\dot\phi_B<0$, the characteristic nonmonotonic behavior familiar
from the LQC case is retained. Representative values are
\begin{align}
\phi_B\simeq1.02:
&\qquad
N_{\rm inf}
=
18.98,\;19.21,\;20.82,
\\
\phi_B\simeq4.02:
&\qquad
N_{\rm inf}
=
50.34,\;49.80,\;46.37,
\\
\phi_B=4.8:
&\qquad
N_{\rm inf}
=
174.03,\;172.62,\;163.56.
\end{align}
The Planck and ACT curves are again very close, while the combined data
produce the largest shift. The dependence on $V_0$ is therefore weak in
the quantum-bounce regime but becomes more visible in the subsequent
inflationary evolution.

Because the PLP potential is even,
\begin{equation}
    V(-\phi)=V(\phi),
\end{equation}
the background equations are invariant under
\begin{equation}
    (\phi,\dot\phi)
    \longrightarrow
    (-\phi,-\dot\phi).
\end{equation}
Therefore the evolution for positive $\phi_B$ and negative $\dot\phi_B$
has a mirror counterpart with negative $\phi_B$ and positive
$\dot\phi_B$, just as in the corresponding LQC analysis.

\subsection{Numerical stability}
\label{subsec:numerical_stability}

The bounce-to-inflation evolution spans many orders of magnitude in
cosmic time, and special care is therefore required in calculating the
number of e-folds. We use a three-stage numerical procedure. Very close
to the bounce, where $H(t_B)=0$, we integrate the canonical mLQC-I
equations in the variables $(b,\phi,\dot\phi,N)$. After a short interval,
we switch to the post-bounce modified Friedmann equation
(\ref{eq:Friedmann_mLQCI}) together with the Klein--Gordon equation.
Finally, once ordinary inflation begins, we use the e-fold number itself
as the independent variable rather than accumulating $\int Hdt$ over the
entire long inflationary interval.

Several independent checks were performed. The explicit post-bounce
$H(\rho)$ relation agrees with the canonical $H(b)$ expression to
typically $10^{-13}$--$10^{-15}$ over the relevant density range, and the
corresponding expressions for $\dot H$ agree at machine precision.
Furthermore, the time-domain evaluation of $N_{\rm inf}$ and the
independent evolution with $N$ as the time variable agree extremely well.
For example, for the Planck normalization and
$\phi_B=-4,\dot\phi_B>0$,
\begin{equation}
N_{\rm inf}^{(t)}
=
48.521206384453,
\qquad
N_{\rm inf}^{(N)}
=
48.521206374659,
\end{equation}
whose difference is below $10^{-8}$. Even for solutions with
$N_{\rm inf}\simeq419$, the fractional discrepancy among independent
implementations remains below $4\times10^{-9}$.

We therefore use the $N$-domain formulation during inflation in all
subsequent calculations. The numerical uncertainty in $N_{\rm inf}$ is
many orders of magnitude below the physical variation caused by the
different observational determinations of $(A_s,n_s)$.

\subsection{Inflationary observables}
\label{subsec:inflation_observables}

Before turning to the reheating epoch, it is useful to summarize the
inflationary predictions of the PLP potential in the $(r,n_s)$ plane.
In the present analysis we work to leading order in the mLQC-I
corrections to the primordial spectra, so that Eq.(\ref{eq:FI_perturb_approx}) holds.
Consequently, at the horizon crossing of the pivot mode, we have \cite{Yogesh:2024iip}
\bqn
\lb{eq:ns_r_mLQCI}
n_s-1
\simeq
-2\left(3\epsilon_V^*-\eta_V^*\right),
\qquad
r_*=16\epsilon_V^* ,
\eqn
where, for the PLP potential with $m=m_{\rm Pl}$,
\bqn
\lb{eq:eps_eta_mLQCI}
\epsilon_V
=
\frac{1}{4\pi(\phi+\phi^3)^2},
\qquad
\eta_V
=
\frac{1-3\phi^2}{4\pi(\phi+\phi^3)^2}.
\eqn
The
quantum-geometric information of mLQC-I enters through the
mapping between the horizon-crossing quantities and the initial data at
the quantum bounce.

The number of e-folds between the horizon crossing of the pivot mode and
the end of inflation is
\bqn
\lb{eq:Nstar_PLP_mLQCI}
N_*
&\equiv&
\ln\left(\frac{a_{\rm end}}{a_*}\right)
\simeq
\frac{8\pi}{m_{\rm Pl}^2}
\int_{\phi_{\rm end}}^{\phi_*}\frac{V}{V_{,\phi}}\,d\phi
=
2\pi\left(\phi_*^2-\phi_{\rm end}^2\right)
+
\pi\left(\phi_*^4-\phi_{\rm end}^4\right),
\eqn
where the second equality is written in the units $m_{\rm Pl}=1$ used in
the numerical analysis.  The value $\phi_{\rm end}$ is determined by
$\epsilon_V(\phi_{\rm end})=1$, which gives
$\phi_{\rm end}\simeq0.26375\,m_{\rm Pl}$.
For illustration, $N_*=50,\;60,$ and $70$ correspond, respectively, to
\bqn
\begin{array}{c|ccc}
N_* & \phi_* & n_s & r_* \\ \hline
50 & 1.76927 & 0.96605 & 0.02384 \\
60 & 1.87061 & 0.97192 & 0.01798 \\
70 & 1.95959 & 0.97608 & 0.01415
\end{array}
\eqn

The three central observational values considered in this work map onto
the PLP curve as
\bqn
\begin{array}{c|ccc}
{\rm Data} & n_s & r_* & N_* \\ \hline
{\rm Planck\ 2018} & 0.9649 & 0.02505 & 48.42 \\
{\rm ACT\ 2025} & 0.9666 & 0.02327 & 50.79 \\
{\rm Combined} & 0.9752 & 0.01494 & 67.61
\end{array}
\eqn
In particular, the larger value of $n_s$ in the combined data set
corresponds to a substantially longer interval between horizon crossing
and the end of inflation.

\begin{figure}[t!]
\centering
\includegraphics[width=0.58\textwidth]{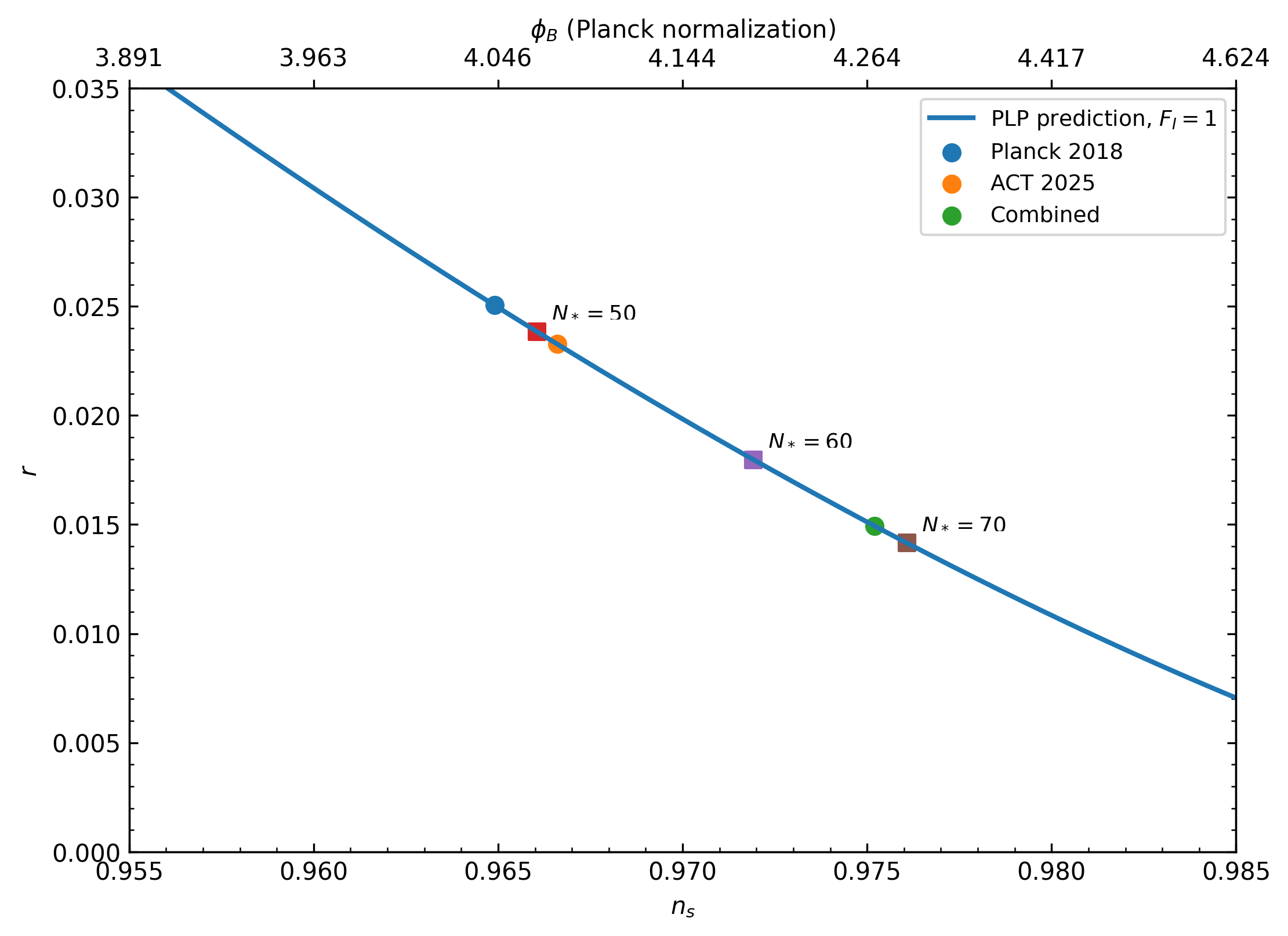}
\caption{The PLP prediction in the $(r,n_s)$ plane in mLQC-I under the
leading-order approximation ${\cal{F}}_{\rm I}(k_*)=1$ and
$\xi_{\rm I}(k_*)=0$.  The square markers denote
$N_*=50,\;60,$ and $70$, while the other markers show the central values
of $n_s$ from Planck 2018, ACT 2025, and the combined
Planck+ACT+DESI DR2 data set mapped onto the PLP trajectory.  The upper
horizontal axis gives the corresponding bounce value $\phi_B$ for the
Planck normalization.  Since the perturbative factor ${\cal{F}}_{\rm I}(k_*)$ is set to unity at
this order, the $r$--$n_s$ trajectory itself is the usual PLP slow-roll
relation; the mLQC-I dependence appears through the mapping to the bounce
data.}
\label{fig:rnsmLQCI}
\end{figure}

Figure~\ref{fig:rnsmLQCI} shows the PLP trajectory in the $(r,n_s)$
plane together with the three central observational values and the
points corresponding to $N_*=50,\;60,$ and $70$.  The upper horizontal
axis shows the associated self-consistent value of $\phi_B$ obtained by
evolving the horizon-crossing data backward to the mLQC-I bounce, using
the Planck normalization.  The dependence of this mapping on the choice
of $A_s$ among the three data sets is very small and does not affect the
conclusions drawn here.

The purpose of Fig.~\ref{fig:rnsmLQCI} is therefore mainly to connect the
observationally inferred values of $(A_s,n_s)$ used in the preceding
background analysis with the inflationary quantities entering the
reheating calculation.  

Taken together, the results of this section show that the genuinely
quantum post-bounce dynamics of mLQC-I is remarkably insensitive to the
observationally inferred normalization $V_0$, whereas the later
inflationary evolution, including $N_{\rm inf}$ and $N_*$, displays a
more visible dependence on the data set.  The $r$--$n_s$ relation itself
remains the standard PLP slow-roll prediction at  leading order
${\cal{F}}_{\rm I}(k_*)=1$, while the mLQC-I information is encoded in the
connection between the horizon-crossing quantities and the bounce data.
These results provide the background input needed for the reheating and
total-expansion analyses of the following sections.

\FloatBarrier
\section{Reheating in mLQC-I}
\label{sec:reheating_mLQCI}

Having determined the inflationary quantities associated with the three
observational data sets, we now turn to the reheating epoch.  As in the
corresponding LQC analysis, we do not specify the detailed microphysics
of reheating.  Instead, we characterize this epoch by an effective
equation-of-state parameter $w_{\rm re}$, the number of reheating
e-folds $N_{\rm re}$, and the temperature $T_{\rm re}$ at the end of
thermalization.

Let $a_{\rm end}$ and $a_{\rm re}$ denote the scale factors at the end
of inflation and at the end of reheating, respectively.  We define
\bqn
\lb{eq:Nre_def_mLQCI}
N_{\rm re}
\equiv
\ln\left(\frac{a_{\rm re}}{a_{\rm end}}\right).
\eqn
Assuming that $w_{\rm re}$ is approximately constant during reheating,
the energy density evolves as
\bqn
\rho_{\rm re}
=
\rho_{\rm end}
e^{-3(1+w_{\rm re})N_{\rm re}} .
\eqn
At the end of slow-roll inflation, $\epsilon_H\simeq\epsilon_V=1$,
which gives $\dot\phi_{\rm end}^{\,2}\simeq V_{\rm end}$ and hence
$\rho_{\rm end}\simeq 3V_{\rm end}/2$.  Using
\bqn
\rho_{\rm re}
=
\frac{\pi^2}{30}g_{\rm re}T_{\rm re}^4,
\eqn
together with entropy conservation after reheating, one obtains
\cite{Yogesh:2024iip}
\bqn
\lb{eq:Nre_mLQCI}
N_{\rm re}
=
-\frac{4}{1-3w_{\rm re}}
\left[
\frac{1}{4}\ln\left(\frac{45}{\pi^2g_{\rm re}}\right)
+
\frac{1}{3}\ln\left(\frac{11g_{\rm re}}{43}\right)
+
\ln\left(\frac{k_*}{a_0T_0}\right)
+
\ln\left(\frac{V_{\rm end}^{1/4}}{H_*}\right)
+
N_*
\right],
\eqn
and
\bqn
\lb{eq:Tre_mLQCI}
T_{\rm re}
=
\left[
\left(\frac{43}{11g_{\rm re}}\right)^{1/3}
\frac{a_0T_0H_*}{k_*}
e^{-N_*}
\left(
\frac{45V_{\rm end}}{\pi^2g_{\rm re}}
\right)^{-\frac{1}{3(1+w_{\rm re})}}
\right]^{
\frac{3(1+w_{\rm re})}{3w_{\rm re}-1}
}.
\eqn
Equations~(\ref{eq:Nre_mLQCI}) and (\ref{eq:Tre_mLQCI}) apply for
$w_{\rm re}\neq1/3$.  The case $w_{\rm re}=1/3$, for which the
reheating energy density redshifts as radiation, must be treated
separately. Here $T_0$ is the present CMB temperature and $g_{\rm re}$ denotes the
number of relativistic degrees of freedom at the end of reheating.  In
the numerical calculations below we take
$g_{\rm re}=226$,
as in Ref.~\cite{Yogesh:2024iip}.  Our conclusions are insensitive to
moderate variations of $g_{\rm re}$ around ${\cal O}(10^2)$.

At the observational pivot scale, and within the leading-order
approximation adopted in Sec.~\ref{sec:background_mLQCI},
\bqn
\lb{eq:Hstar_reheating_mLQCI}
H_*
=
\left[
\frac{\pi r_* A_s}{16{\cal{F}}_{\rm I}(k_*)}
\right]^{1/2}m_{\rm Pl}
\simeq
\sqrt{\pi\epsilon_V^*A_s}\,m_{\rm Pl},
\eqn
where ${\cal{F}}_{\rm I}(k_*)=1$ and $r_*=16\epsilon_V^*$.  The end value
$V_{\rm end}$ is evaluated at $\phi_{\rm end}$ determined by
$\epsilon_V(\phi_{\rm end})=1$.

\begin{figure*}[t!]
\centering
\includegraphics[width=0.95\textwidth]{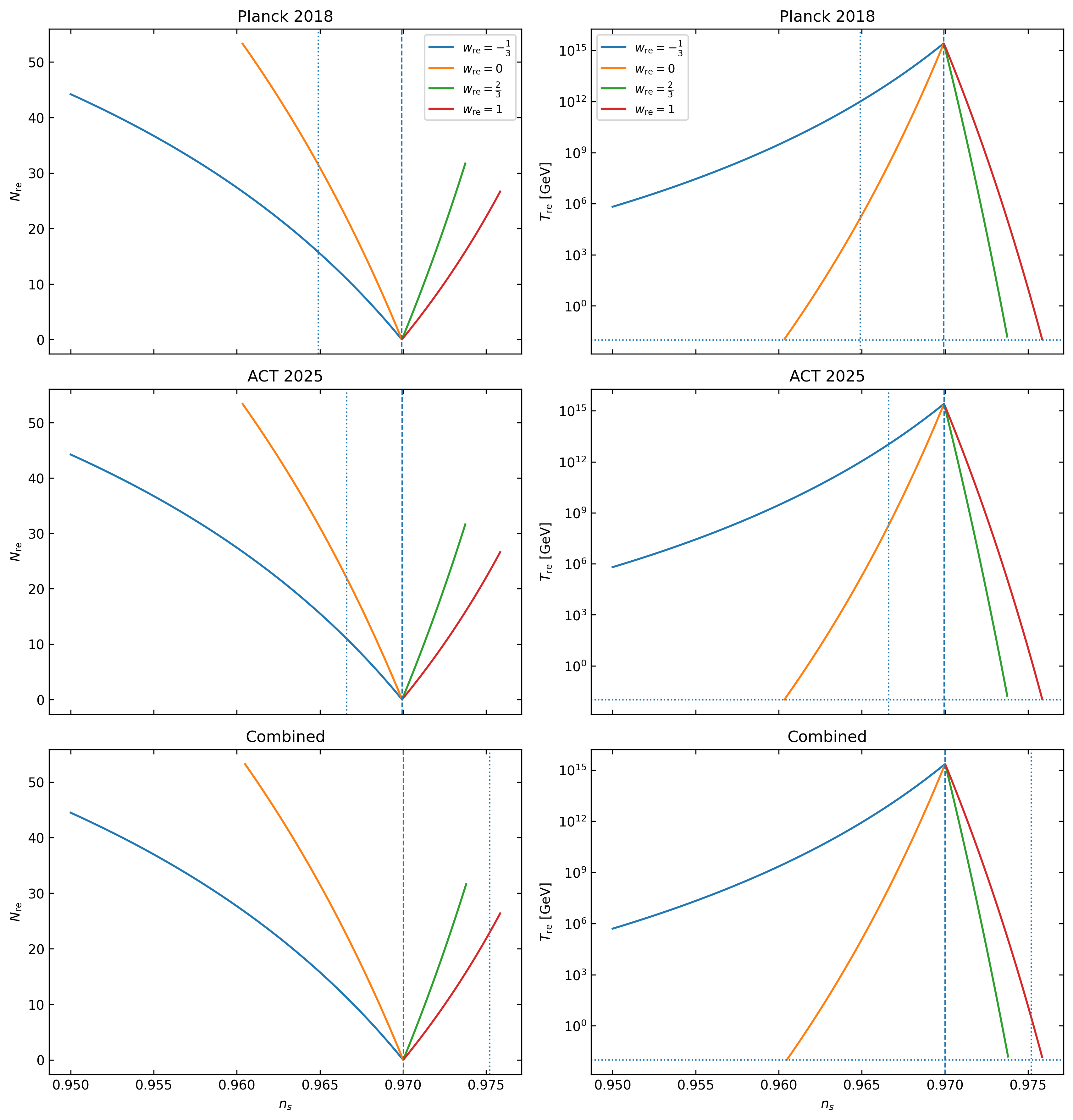}
\caption{Reheating e-folds $N_{\rm re}$ (left column) and reheating
temperature $T_{\rm re}$ (right column) as functions of $n_s$ for the
Planck 2018, ACT 2025, and combined Planck+ACT+DESI DR2 normalizations.
For each data set, $A_s$ and $V_0$ are fixed at the values inferred from
its central observational pair, while $n_s$ is varied.  The curves
correspond to $w_{\rm re}=-1/3,\,0,\,2/3,$ and $1$.  Only the portions
satisfying $N_{\rm re}\geq0$ and
$10\,{\rm MeV}\leq T_{\rm re}\leq T_{\rm re}^{\rm inst}$ are shown.
The vertical dotted lines indicate the central observational values of
$n_s$, while the vertical dashed lines mark instantaneous reheating.}
\label{fig:reheating_mLQCI}
\end{figure*}

Figure~\ref{fig:reheating_mLQCI} shows $N_{\rm re}$ and
$T_{\rm re}$ as functions of $n_s$ for the Planck 2018, ACT 2025, and
combined Planck+ACT+DESI DR2 normalizations.  For each panel, $A_s$ and
$V_0$ are fixed by the central observational pair of the corresponding
data set, while $n_s$ is varied in order to display the reheating
dependence in the same manner as in the original LQC analysis.  We
consider four representative values
\bqn
w_{\rm re}
=
-\frac{1}{3},\quad
0,\quad
\frac{2}{3},\quad
1 .
\eqn
Only portions of the curves satisfying
\bqn
N_{\rm re}\geq0,
\qquad
10\,{\rm MeV}
\leq T_{\rm re}
\leq T_{\rm re}^{\rm inst},
\label{eq:physical_reheating_region}
\eqn
are retained, where $T_{\rm re}^{\rm inst}$ denotes the instantaneous
reheating scale.

For all three normalizations, the instantaneous-reheating point lies
very close to
$n_s^{\rm inst}\simeq0.970$.
More precisely, we obtain
\bqn
\begin{array}{c|cc}
{\rm Data}
& n_s^{\rm inst}
& T_{\rm re}^{\rm inst}\ ({\rm GeV})
\\ \hline
{\rm Planck\ 2018}
&0.9699345
&2.63\times10^{15}
\\
{\rm ACT\ 2025}
&0.9699469
&2.58\times10^{15}
\\
{\rm Combined}
&0.9700142
&2.29\times10^{15}
\end{array}
\label{eq:instantaneous_reheating_values}
\eqn

At the central Planck value $n_s=0.9649$, the two illustrative cases
$w_{\rm re}=-1/3$ and $w_{\rm re}=0$ give
\bqn
\begin{array}{c|cc}
w_{\rm re}
&N_{\rm re}
&T_{\rm re}\ ({\rm GeV})
\\ \hline
-1/3
&15.76
&9.93\times10^{11}
\\
0
&31.52
&1.42\times10^{5}
\end{array}
\label{eq:Planck_reheating_values}
\eqn
whereas the cases $w_{\rm re}>1/3$ lead to $N_{\rm re}<0$ at the
central Planck value and are therefore not physically realized within
this effective reheating description.

Similarly, for ACT 2025, with $n_s=0.9666$, we find
\bqn
\begin{array}{c|cc}
w_{\rm re}
&N_{\rm re}
&T_{\rm re}\ ({\rm GeV})
\\ \hline
-1/3
&11.00
&1.05\times10^{13}
\\
0
&22.00
&1.76\times10^{8}
\end{array}
\label{eq:ACT_reheating_values}
\eqn
Thus, Planck and ACT favor very similar qualitative reheating histories
within the four representative equations of state considered here.

The combined data lead to a qualitatively different situation because
their central value,
$n_s=0.9752$,
lies above $n_s^{\rm inst}$.  In this case the branches with
$w_{\rm re}<1/3$ have negative $N_{\rm re}$.  For the two cases
$w_{\rm re}>1/3$ we obtain
\bqn
\begin{array}{c|cc}
w_{\rm re}
&N_{\rm re}
&T_{\rm re}\ ({\rm GeV})
\\ \hline
2/3
&45.68
&3.64\times10^{-10}
\\
1
&22.84
&3.03
\end{array}
\label{eq:combined_reheating_values}
\eqn
The $w_{\rm re}=2/3$ case falls far below the BBN temperature and is
therefore excluded by Eq.~(\ref{eq:physical_reheating_region}), while
the $w_{\rm re}=1$ case remains compatible with the BBN bound.  We
stress that these statements refer only to the four representative
constant-$w_{\rm re}$ cases considered here and should not be interpreted
as a determination of the microscopic reheating equation of state.

The results demonstrate that the reheating formalism can be implemented
consistently in mLQC-I once the inflationary normalization is fixed by
observations.  at  leading order, with ${\cal{F}}_{\rm I}(k_*)=1$, the explicit
reheating formulas retain their standard form; the role of mLQC-I enters
through the background history and, most importantly, through the
connection between the pivot scale and the quantum bounce.  This
connection is investigated in the next section.

\FloatBarrier
\section{Reheating and the total expansion history}
\label{sec:total_expansion_mLQCI}

We now consider the total amount of expansion from the mLQC-I bounce to
the present epoch.  With $a_B=1$, the total number of e-folds is
\bqn
\lb{eq:NT_def_mLQCI}
N_T
&\equiv&
\ln\left(\frac{a_0}{a_B}\right)
=
\ln\left(\frac{a_*}{a_B}\right)
+
N_*
+
N_{\rm re}
+
\ln\left(\frac{a_0}{a_{\rm re}}\right).
\eqn
Entropy conservation from the end of reheating to the present gives
\bqn
\lb{eq:a0_are_mLQCI}
\frac{a_0}{a_{\rm re}}
=
\left(\frac{11g_{\rm re}}{43}\right)^{1/3}
\frac{T_{\rm re}}{T_0}.
\eqn
Substituting Eqs.~(\ref{eq:Nre_mLQCI}) and
(\ref{eq:Tre_mLQCI}) into Eqs.~(\ref{eq:NT_def_mLQCI}) and
(\ref{eq:a0_are_mLQCI}), the explicit dependence on
$N_{\rm re}$ and $w_{\rm re}$ cancels.  Using the horizon-crossing
condition
\bqn
k_*=a_*H_*=a_0 k_*^{\rm ob},
\qquad
k_*^{\rm ob}=0.05\ {\rm Mpc}^{-1},
\eqn
one obtains the particularly transparent expression
\bqn
\lb{eq:NT_simple_mLQCI}
N_T
=
\ln\left(\frac{a_*}{a_B}\right)
+
\ln\left(\frac{H_*}{k_*^{\rm ob}}\right).
\eqn
Equation~(\ref{eq:NT_simple_mLQCI}) is useful both conceptually and
numerically.  It shows that, once the observed pivot scale is fixed, the
total expansion is determined by the horizon-crossing quantities and by
the mLQC-I evolution connecting $t_*$ to the bounce.

For a given observational pair $(A_s,n_s)$, we first solve the
slow-roll equations to determine $\phi_*$, $H_*$, $V_0$, and
$\dot\phi_*$.  These data are then evolved backward using the
post-bounce mLQC-I equations until the quantum bounce is reached.  The
rescaling symmetry is used to impose $a_B=1$, which determines
$a_*/a_B$.  This procedure is the mLQC-I counterpart of the nonlocal
normalization condition used in \cite{Yogesh:2024iip} for the LQC analysis.

For the three central observational data sets we obtain
\bqn
\begin{array}{c|ccc}
{\rm Data}
&\phi_B
&\ln(a_*/a_B)
&N_T
\\ \hline
{\rm Planck\ 2018}
& 4.04406
& 7.18714
&129.42869
\\
{\rm ACT\ 2025}
& 4.07467
& 7.22611
&129.43524
\\
{\rm Combined}
& 4.26820
& 7.46778
&129.45971
\end{array}
\label{eq:NT_central_mLQCI}
\eqn
The entries in Eq.~(\ref{eq:NT_central_mLQCI}) are shown for the branch
$\phi_B>0,\dot\phi_B<0$; by the symmetry of the PLP potential they are
equivalent to the corresponding
$\phi_B<0,\dot\phi_B>0$ solutions.

\begin{figure}[t!]
\centering
\includegraphics[width=0.62\textwidth]{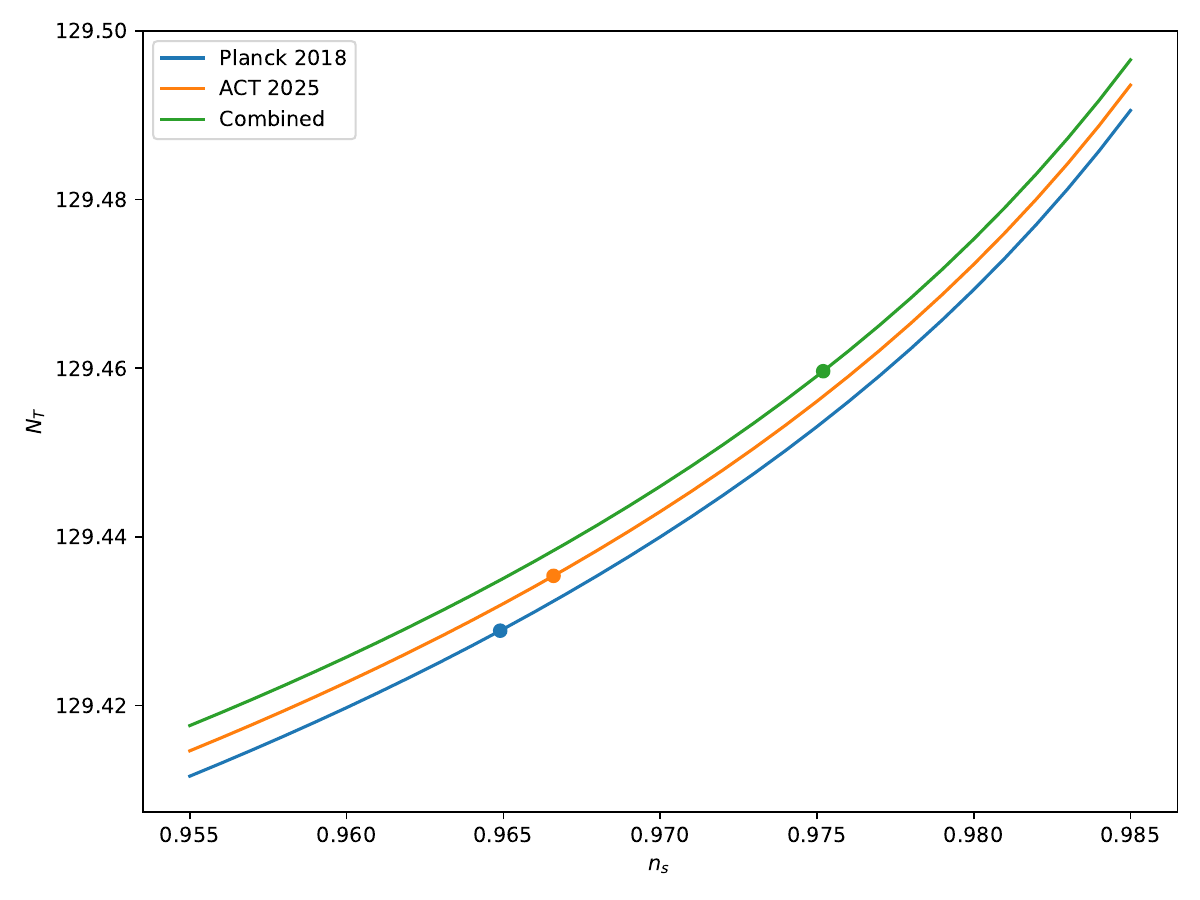}
\caption{Total number of e-folds $N_T$ from the mLQC-I bounce to the
present as a function of $n_s$.  For each curve, $A_s$ is held fixed at
the corresponding observational value, while $V_0(n_s)$ is
re-determined self-consistently from the scalar amplitude and the
slow-roll Friedmann equation.  The markers denote the Planck 2018,
ACT 2025, and combined Planck+ACT+DESI DR2 central values.  Over the
displayed interval $0.955\leq n_s\leq0.985$, the curves remain close to
$129.4$ and their smallest displayed value occurs at the lower boundary,
giving $N_T\gtrsim129.4$ within this interval, with a characteristic
total expansion scale $N_T\sim130$.}
\label{fig:NT_mLQCI}
\end{figure}

Figure~\ref{fig:NT_mLQCI} displays $N_T$ as a function of $n_s$.
For the continuous curves, $A_s$ is held fixed at the value of each
observational data set while $V_0(n_s)$ is re-determined at every
$n_s$ so that the scalar-amplitude and slow-roll Friedmann relations
remain mutually consistent.  The markers denote the actual central
observational pairs.  Over the illustrative interval
\bqn
0.955\leq n_s\leq0.985,
\eqn
the curves are monotonic and their smallest displayed value occurs at
the lower boundary, $n_s=0.955$.  Numerically, we find the lower bound
$N_T\gtrsim129.41$,
or, more conservatively,
\bqn
N_T\gtrsim129.
\eqn
Thus the characteristic total expansion remains of order
$N_T\sim130$.
Here the bound refers specifically to the displayed interval in $n_s$;
it should not be interpreted as a global lower bound outside the range
considered.

It is useful to compare this result directly with the corresponding LQC
background evolution under the same leading-order assumptions.  Using
the same three observational inputs, we obtain
\bqn
\begin{array}{c|ccc}
{\rm Data}
&N_T^{\rm LQC}
&N_T^{\rm mLQC-I}
&N_T^{\rm mLQC-I}-N_T^{\rm LQC}
\\ \hline
{\rm Planck\ 2018}
&129.66907
&129.42869
&-0.24038
\\
{\rm ACT\ 2025}
&129.67557
&129.43524
&-0.24032
\\
{\rm Combined}
&129.69983
&129.45971
&-0.24012
\end{array}.
\label{eq:LQC_mLQCI_comparison}
\eqn
Thus, although both LQC and mLQC-I lead to the same characteristic scale
$N_T\sim130$, the modified background dynamics of mLQC-I produces a
small but systematic shift,
\bqn
\lb{eq:DeltaNT}
\Delta N_T
\equiv
N_T^{\rm mLQC-I}-N_T^{\rm LQC}
\simeq
-0.24.
\eqn
This shift is essentially independent of which of the three
observational normalizations is used.  It is also much larger than the
numerical integration uncertainty: tightening the integration tolerances
changes the central values of $N_T$ by only about
$5\times10^{-9}$--$6\times10^{-9}$.

The result in Eq.~(\ref{eq:DeltaNT}) provides a direct quantitative
measure of the difference between the LQC and mLQC-I post-bounce
histories for the same inflationary potential and the same observational
normalization.  At the same time, the result
$N_T\gtrsim129$ over the displayed observationally relevant interval
demonstrates that the principal conclusion of the LQC reheating analysis
is robust under the replacement of LQC by mLQC-I, at least within the
leading-order treatment ${\cal{F}}_{\rm I}(k_*)=1$ adopted here.

\FloatBarrier
\section{Sensitivity to the Inflationary Potential}
\label{sec:potential_dependence}

The results of Secs.~II--IV were obtained for the power-law plateau
(PLP) potential.  It is therefore important to determine which of our
main conclusions are specific to that choice and which remain stable
when the inflationary potential is changed.  In this section we consider
two additional examples: the Starobinsky potential
\cite{Starobinsky:1980te} and the polynomial chaotic potential used in
Refs.~\cite{Brown:2025hcb,Fier:2026xko}.  In both cases we use the same three
observational data sets as in Sec.~II and keep the leading-order
approximation
${\cal{F}}_{\rm I}(k_*)=1$.
The general slow-roll relations, the reheating formulas, and the
expression for the total expansion have already been derived in the
preceding sections and will not be repeated here.  Instead, for each
potential we determine its parameters from the observed pair
$(A_s,n_s)$ and then apply Eqs.~(\ref{eq:Nre_mLQCI}),
(\ref{eq:Tre_mLQCI}), and (\ref{eq:NT_simple_mLQCI}).

\subsection{Starobinsky potential}
\label{subsec:Starobinsky}

We first consider the Starobinsky potential
\begin{equation}
V(\phi)
=
V_0\left(1-e^{-\alpha\phi}\right)^2,
\qquad
\alpha=\sqrt{\frac{16\pi}{3}},
\label{eq:Starobinsky_potential}
\end{equation}
where $m_{\rm Pl}=1$ is understood in this subsection.  Restoring
dimensions, $V_0$ carries units of $m_{\rm Pl}^4$ and the exponent
contains $\phi/m_{\rm Pl}$.  The corresponding potential slow-roll
parameters are
\begin{align}
\epsilon_V
&=
\frac{4}{3}
\left(
\frac{e^{-\alpha\phi}}
     {1-e^{-\alpha\phi}}
\right)^2,
\label{eq:Starobinsky_epsilon}
\\
\eta_V
&=
\frac{4}{3}
\frac{-e^{-\alpha\phi}+2e^{-2\alpha\phi}}
     {\left(1-e^{-\alpha\phi}\right)^2}.
\label{eq:Starobinsky_eta}
\end{align}

For each observed value of $n_s$, Eqs.~(\ref{eq:Starobinsky_epsilon})
and (\ref{eq:Starobinsky_eta}), together with the general slow-roll
relation (\ref{eq:nsr_slowroll}), determine $\phi_*$.  The scalar
amplitude then fixes $H_*$ through Eq.~(\ref{eq:Hstar_As}), while
Eq.~(\ref{eq:Hstar_V}) determines the normalization $V_0$.  Finally,
$N_*$ is obtained from the slow-roll integral in
Eq.~(\ref{eq:Nstar_PLP_mLQCI}), with the Starobinsky potential
(\ref{eq:Starobinsky_potential}); the end of inflation is fixed by
$\epsilon_V(\phi_{\rm end})=1$, which gives
$\phi_{\rm end}\simeq0.18754$ in the present units.
The resulting central values are summarized in
Table~\ref{tab:Starobinsky_parameters}, which  shows that the Planck and ACT
normalizations are again close to each other, whereas the combined data
prefer a considerably smaller potential scale.  More importantly, the
larger central value of $n_s$ in the combined data corresponds to a much
longer interval between horizon crossing and the end of inflation:
$N_*$ increases from $54.88$ for Planck to $78.30$ for the combined
data.  This difference has an immediate impact on the reheating
constraints.

\begin{table}[!ht]
\centering
\caption{Starobinsky parameters inferred from the three observational
data sets, assuming ${\cal{F}}_{\rm I}(k_*)=1$.  The entries for $V_0$ are in
units of $m_{\rm Pl}^4$.}
\label{tab:Starobinsky_parameters}
\begin{tabular}{lcccc}
\hline\hline
Data set
& $\phi_*/m_{\rm Pl}$
& $r_*$
& $V_0/m_{\rm Pl}^4$
& $N_*$
\\
\hline
Planck 2018
& 1.06722
& 0.003513
& $1.7730\times10^{-13}$
& 54.88
\\
ACT 2025
& 1.07892
& 0.003189
& $1.6218\times10^{-13}$
& 57.74
\\
Combined
& 1.14942
& 0.001780
& $9.0764\times10^{-14}$
& 78.30
\\
\hline\hline
\end{tabular}
\end{table}

\begin{figure*}[!ht]
\centering
\includegraphics[width=0.95\textwidth]{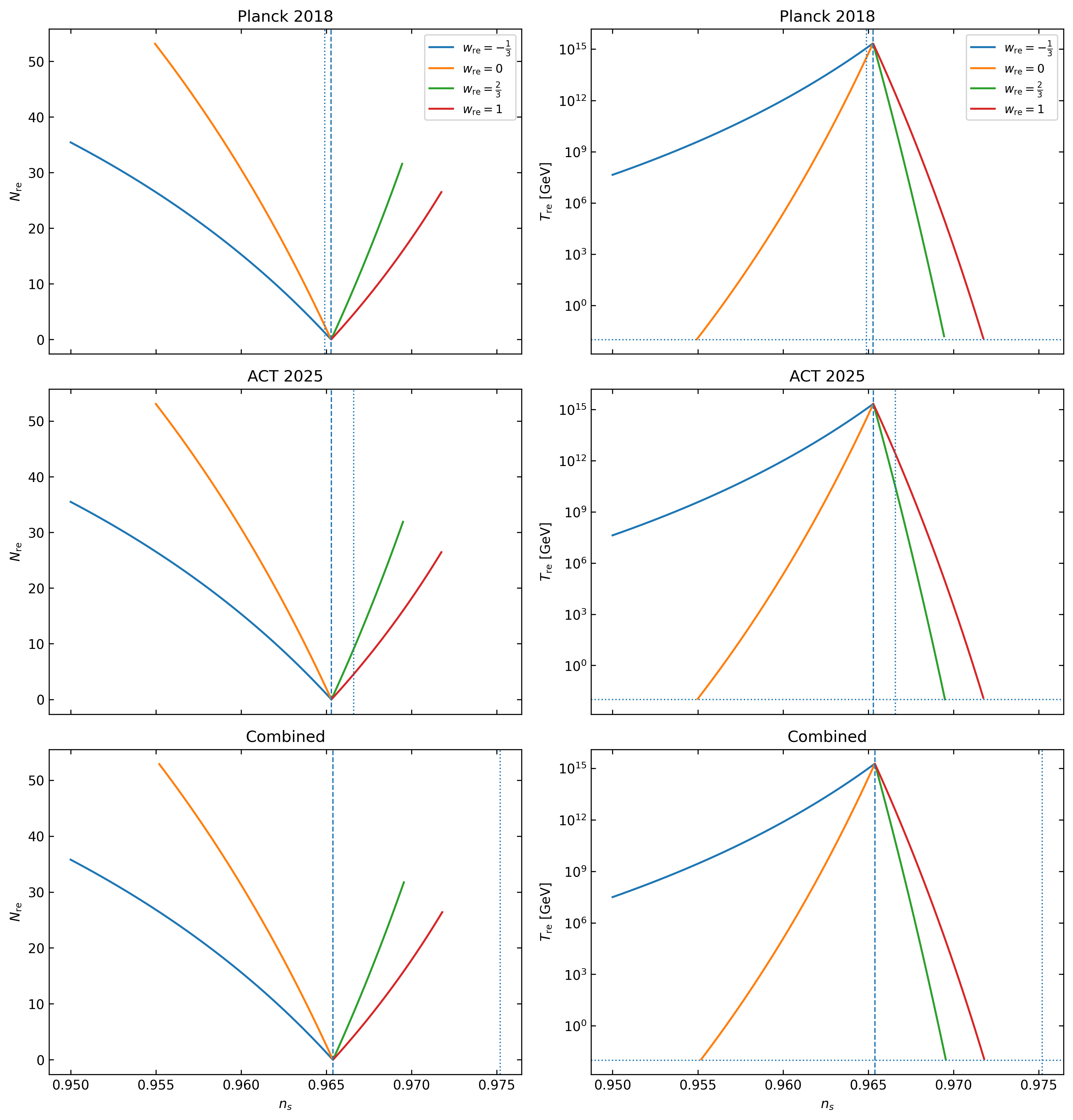}
\caption{Reheating e-folds $N_{\rm re}$ and reheating temperature
$T_{\rm re}$ for the Starobinsky potential.  The conventions are the
same as in Fig.~\ref{fig:reheating_mLQCI}.  The instantaneous-reheating
point is shifted from $n_s\simeq0.970$ for the PLP potential to
$n_s\simeq0.9653$.}
\label{fig:Starobinsky_reheating}
\end{figure*}

To construct the reheating curves, we follow the same prescription used
for Fig.~\ref{fig:reheating_mLQCI}: for each observational data set,
$A_s$ and the value of $V_0$ inferred from its central pair $(A_s,n_s)$
are held fixed while $n_s$ is varied.  The resulting $N_{\rm re}$ and
$T_{\rm re}$ are shown in Fig.~\ref{fig:Starobinsky_reheating}.

A comparison of Figs.~\ref{fig:reheating_mLQCI} and
\ref{fig:Starobinsky_reheating} shows that the reheating constraints are
substantially more sensitive to the shape of the inflationary potential
than the immediate post-bounce dynamics.  For the Starobinsky model,
the instantaneous-reheating point is
\begin{equation}
n_s^{\rm inst}\simeq0.9653,
\label{eq:Star_nsinst}
\end{equation}
with the more precise values
\begin{equation}
n_s^{\rm inst}
=
0.965282,\quad
0.965298,\quad
0.965388,
\label{eq:Star_nsinst_values}
\end{equation}
and
\begin{equation}
T_{\rm re}^{\rm inst}
=
2.19\times10^{15},\quad
2.14\times10^{15},\quad
1.85\times10^{15}\ {\rm GeV},
\label{eq:Star_Tinst_values}
\end{equation}
for Planck, ACT, and the combined data, respectively.

At the Planck central value, the representative cases
$w_{\rm re}=-1/3$ and $w_{\rm re}=0$ remain viable and give
\begin{align}
N_{\rm re}
&=
1.26,\qquad 2.52,
\\
T_{\rm re}
&=
1.16\times10^{15},\qquad
3.30\times10^{14}\ {\rm GeV},
\end{align}
respectively.  For ACT, however, the central value
$n_s=0.9666$ lies above the instantaneous-reheating point.  Among the
four representative equations of state considered in Sec.~III, the
physically relevant branches are therefore those with
$w_{\rm re}>1/3$:
\begin{align}
w_{\rm re}=\frac{2}{3}:
&\quad
N_{\rm re}=9.03,
\qquad
T_{\rm re}=2.68\times10^{10}\ {\rm GeV},
\\
w_{\rm re}=1:
&\quad
N_{\rm re}=4.51,
\qquad
T_{\rm re}=2.45\times10^{12}\ {\rm GeV}.
\end{align}
At the combined central value the two $w_{\rm re}>1/3$ branches have
positive $N_{\rm re}$, but their reheating temperatures fall well below
the BBN scale.  Hence none of the four representative constant
$w_{\rm re}$ cases considered here satisfies the BBN bound at the
combined central value.  This is qualitatively different from the PLP
case, for which the $w_{\rm re}=1$ branch remains above the BBN
temperature.

\begin{figure}[!ht]
\centering
\includegraphics[width=0.70\textwidth]{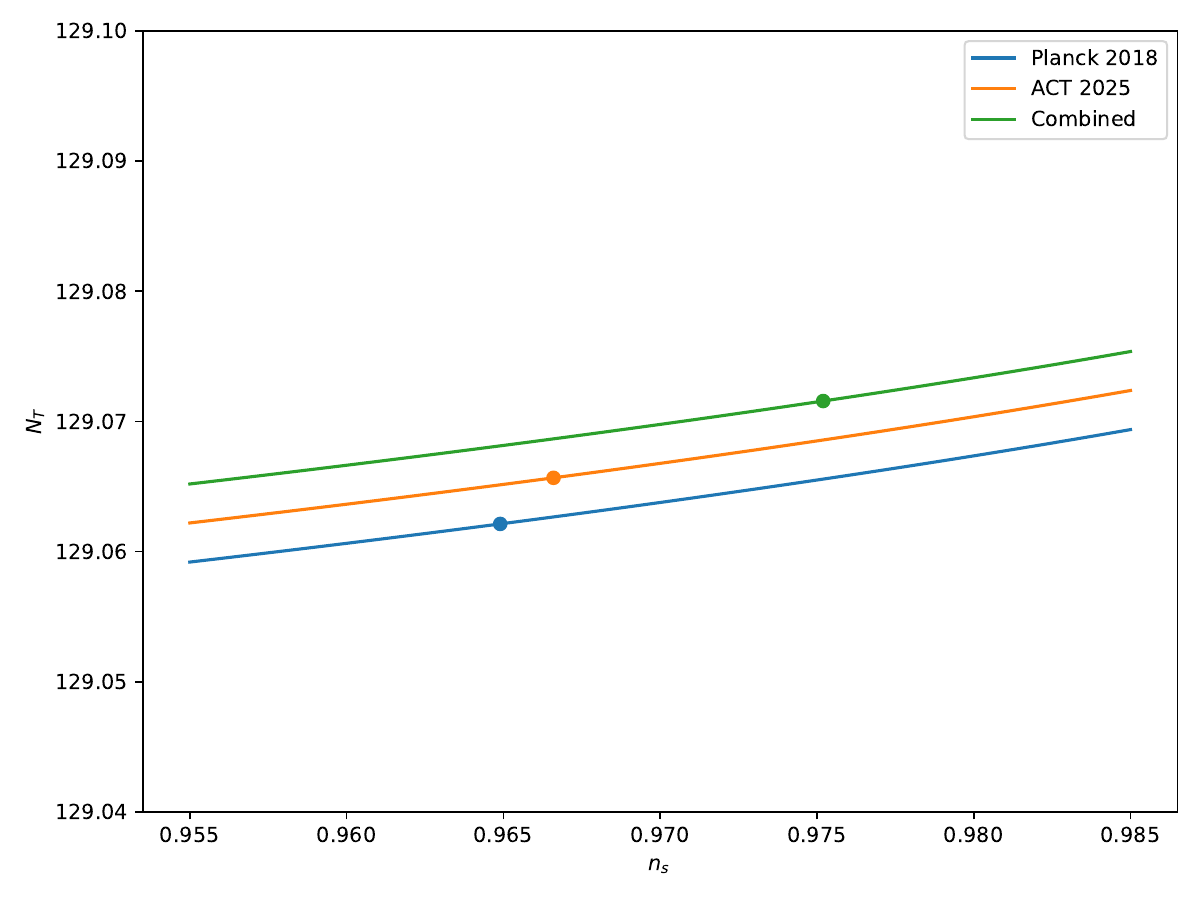}
\caption{Total number of e-folds $N_T$ for the Starobinsky potential.
For the continuous curves $A_s$ is fixed while $V_0(n_s)$ is
re-determined self-consistently at each $n_s$.  The absolute value of
$N_T$ differs modestly from the PLP result, while the characteristic
scale remains $N_T\sim130$.}
\label{fig:Starobinsky_NT}
\end{figure}

We next examine the total expansion.  As in Fig.~\ref{fig:NT_mLQCI},
the continuous $N_T(n_s)$ curves are constructed by keeping $A_s$ fixed
while re-determining $V_0(n_s)$ at every value of $n_s$ so that the
scalar amplitude and slow-roll Friedmann equation remain mutually
consistent.  The three central values are listed in
Table~\ref{tab:Starobinsky_NT}, and the corresponding curves are shown
in Fig.~\ref{fig:Starobinsky_NT}.
Table~\ref{tab:Starobinsky_NT} and Fig.~\ref{fig:Starobinsky_NT}
show that changing from the PLP to the Starobinsky potential shifts the
absolute value of the total expansion from approximately $129.4$ to
approximately $129.06$ e-folds at the central observational points.
This change is visible but modest compared with the strong shift in the
reheating constraints.  More strikingly, the difference
$\Delta N_T
\equiv
N_T^{\rm mLQC-I}-N_T^{\rm LQC}$
%
remains essentially unchanged:
\begin{equation}
\Delta N_T\simeq-0.24019,
\label{eq:DeltaNT_Star}
\end{equation}
for all three data sets.

\begin{table}[!ht]
\centering
\caption{Total expansion for the Starobinsky potential and the
corresponding LQC comparison.}
\label{tab:Starobinsky_NT}
\begin{tabular}{lccc}
\hline\hline
Data set
& $N_T^{\rm mLQC-I}$
& $N_T^{\rm LQC}$
& $\Delta N_T$
\\
\hline
Planck 2018
& 129.06212
& 129.30232
& $-0.24019$
\\
ACT 2025
& 129.06566
& 129.30586
& $-0.24019$
\\
Combined
& 129.07156
& 129.31176
& $-0.24019$
\\
\hline\hline
\end{tabular}
\end{table}

\begin{table}[!ht]
\centering
\caption{Total expansion for the Starobinsky potential and the
corresponding LQC comparison.}
\label{tab:Starobinsky_NT}
\begin{tabular}{lccc}
\hline\hline
Data set
& $N_T^{\rm mLQC-I}$
& $N_T^{\rm LQC}$
& $\Delta N_T$
\\
\hline
Planck 2018
& 129.06212
& 129.30232
& $-0.24019$
\\
ACT 2025
& 129.06566
& 129.30586
& $-0.24019$
\\
Combined
& 129.07156
& 129.31176
& $-0.24019$
\\
\hline\hline
\end{tabular}
\end{table}

\FloatBarrier
\subsection{Polynomial chaotic potential}
\label{subsec:polynomial_chaotic}

We now turn to the polynomial chaotic potential
\begin{equation}
V(\phi)
=
\frac12 m^2\phi^2
\left(
1-\alpha_1\phi+\alpha_2\phi^2
\right)^2,
\label{eq:poly_potential}
\end{equation}
with
$\alpha_1=0.14,
\;
\alpha_2=6.644\times10^{-3}$,
as in Refs.~\cite{Brown:2025hcb,Fier:2026xko}.  Again $m_{\rm Pl}=1$ is used in the
numerical calculations.

The parameters are obtained by substituting
Eq.~(\ref{eq:poly_potential}) into the slow-roll definitions used in
Sec.~II and solving Eq.~(\ref{eq:nsr_slowroll}) for a given observed
$n_s$.  Unlike the PLP and Starobinsky cases, this equation possesses
two relevant roots: one with $\phi_*>0$ and another with $\phi_*<0$.
For each root, $H_*$ is determined from Eq.~(\ref{eq:Hstar_As}), while
Eq.~(\ref{eq:Hstar_V}) fixes the corresponding value of the mass
parameter $m$.  The e-fold number $N_*$ follows from the same slow-roll
integral used in Eq.~(\ref{eq:Nstar_PLP_mLQCI}).  Thus a single
observational pair $(A_s,n_s)$ leads to two distinct inflationary
branches rather than to a unique mass parameter.
The resulting values are collected in Table~\ref{tab:poly_branches}.

\begin{table*}[!ht]
\centering
\caption{Two observational branches of the polynomial chaotic
potential for $\alpha_1=0.14$ and
$\alpha_2=6.644\times10^{-3}$.  Masses and Hubble parameters are shown
in units of $m_{\rm Pl}$.}
\label{tab:poly_branches}
\begin{tabular}{lcccccc}
\hline\hline
Data set
& branch
& $\phi_*/m_{\rm Pl}$
& $m/m_{\rm Pl}$
& $H_*/m_{\rm Pl}$
& $N_*$
& $N_T^{\rm mLQC-I}$
\\
\hline
Planck 2018
& $\phi_*>0$
& 2.52120
& $1.5594\times10^{-6}$
& $5.5463\times10^{-6}$
& 53.89
& 129.7283
\\
Planck 2018
& $\phi_*<0$
& $-3.78698$
& $6.9303\times10^{-7}$
& $8.7310\times10^{-6}$
& 68.72
& 131.2002
\\
ACT 2025
& $\phi_*>0$
& 2.57517
& $1.4918\times10^{-6}$
& $5.3745\times10^{-6}$
& 56.75
& 129.7316
\\
ACT 2025
& $\phi_*<0$
& $-3.90482$
& $6.5086\times10^{-7}$
& $8.5721\times10^{-6}$
& 72.65
& 131.3511
\\
Combined
& $\phi_*>0$
& 2.92033
& $1.1116\times10^{-6}$
& $4.3042\times10^{-6}$
& 77.60
& 129.7323
\\
Combined
& $\phi_*<0$
& $-4.70488$
& $4.3162\times10^{-7}$
& $7.5050\times10^{-6}$
& 101.65
& 131.2045
\\
\hline\hline
\end{tabular}
\end{table*}

Table~\ref{tab:poly_branches} makes clear that the two solutions are
physically distinct.  The $\phi_*>0$ branch has the larger mass,
$m\sim10^{-6}m_{\rm Pl}$, and yields $N_*\simeq54$--$78$ for the three
data sets.  The $\phi_*<0$ branch has a smaller mass and considerably
larger values of $N_*$, ranging from about $69$ to $102$.  The
two-branch degeneracy is therefore not merely a degeneracy in the fitted
mass parameter; it corresponds to different inflationary histories and,
as we now show, different reheating predictions.

\begin{figure*}[!ht]
\centering
\includegraphics[width=0.95\textwidth]{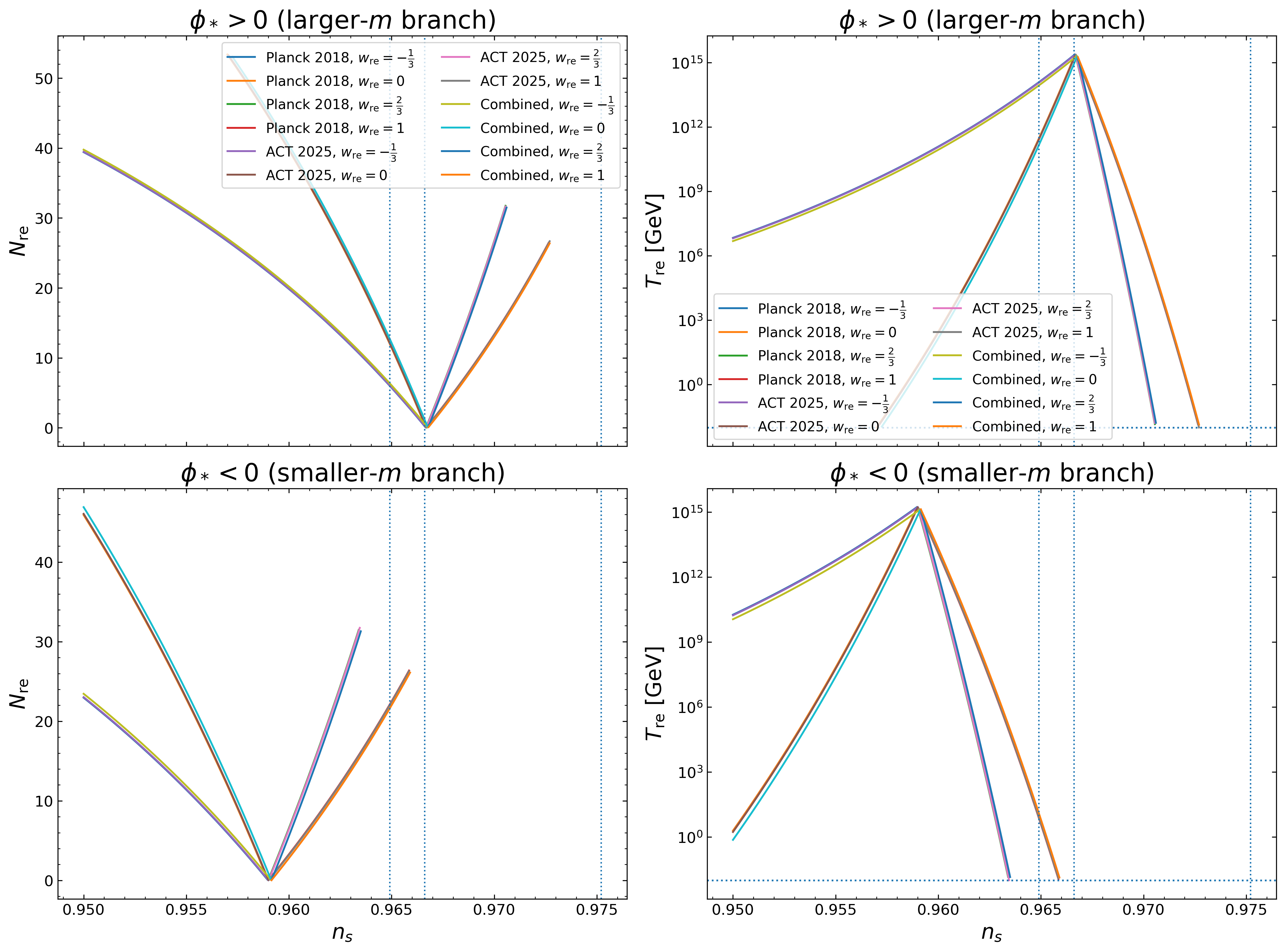}
\caption{Reheating predictions for the polynomial chaotic potential.
The upper row corresponds to the larger-$m$ solution with $\phi_*>0$,
while the lower row corresponds to the smaller-$m$ solution with
$\phi_*<0$.  For each branch the mass inferred from the central
observational pair is held fixed while $n_s$ is varied.}
\label{fig:poly_reheating}
\end{figure*}

For the reheating curves in Fig.~\ref{fig:poly_reheating}, the
procedure is the same as in Figs.~\ref{fig:reheating_mLQCI} and
\ref{fig:Starobinsky_reheating}.  For each data set and each branch,
$A_s$ and the mass $m$ determined from the corresponding central
observational pair are kept fixed while $n_s$ is varied.

Figure~\ref{fig:poly_reheating} displays a pronounced branch
dependence.  The instantaneous-reheating point is approximately
\begin{equation}
n_s^{\rm inst}\simeq0.9667
\qquad
(\phi_*>0),
\label{eq:poly_nsinst_plus}
\end{equation}
whereas
\begin{equation}
n_s^{\rm inst}\simeq0.9590
\qquad
(\phi_*<0).
\label{eq:poly_nsinst_minus}
\end{equation}
For the Planck central value, the larger-$m$ branch allows both
$w_{\rm re}=-1/3$ and $w_{\rm re}=0$, with
\begin{align}
N_{\rm re}
&=
6.04,\qquad12.07,
\\
T_{\rm re}
&=
1.21\times10^{14},\qquad
2.88\times10^{11}\ {\rm GeV},
\end{align}
respectively.  On the smaller-$m$ branch these two cases instead have
negative $N_{\rm re}$.  Among the four representative equations of
state considered here, only the $w_{\rm re}=1$ case remains above the
BBN temperature, for which
\begin{equation}
N_{\rm re}=22.06,
\qquad
T_{\rm re}\simeq7.56\ {\rm GeV}.
\end{equation}

For ACT, the larger-$m$ branch lies very close to instantaneous
reheating:
\begin{align}
w_{\rm re}=-\frac13:
&\quad
N_{\rm re}=0.314,
\qquad
T_{\rm re}=2.06\times10^{15}\ {\rm GeV},
\\
w_{\rm re}=0:
&\quad
N_{\rm re}=0.628,
\qquad
T_{\rm re}=1.51\times10^{15}\ {\rm GeV}.
\end{align}
By contrast, on the smaller-$m$ branch none of the four representative
equations of state satisfies both $N_{\rm re}\geq0$ and the BBN
temperature bound.  At the combined central value, neither branch
satisfies the BBN bound for the four representative constant
$w_{\rm re}$ choices.  Thus the polynomial model demonstrates that the
reheating constraints can be sensitive not only to the functional form
of the inflaton potential but also to which of two observationally
allowed branches is realized.

\begin{figure}[!ht]
\centering
\includegraphics[width=0.72\textwidth]{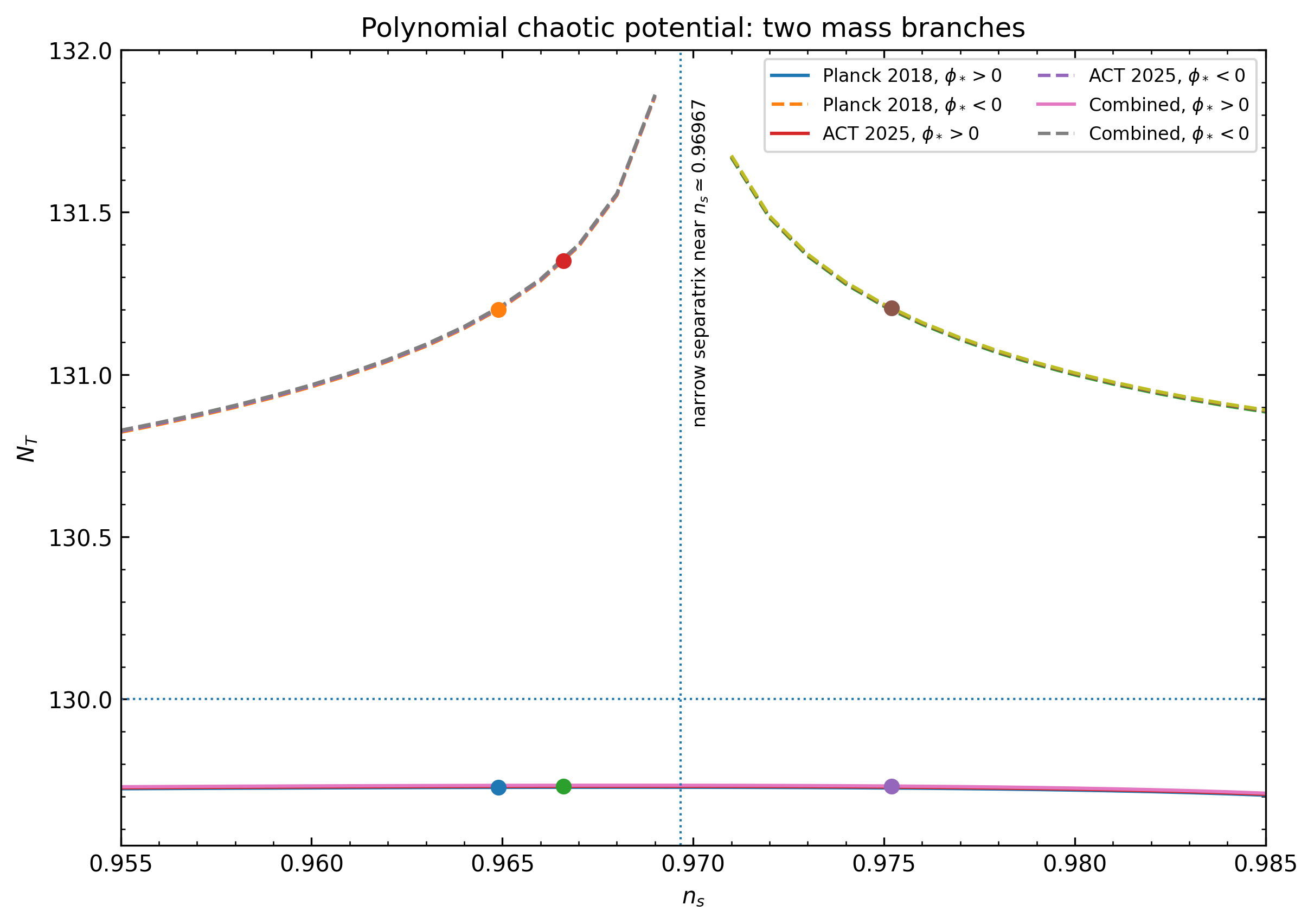}
\caption{Total number of e-folds for the two observational branches of
the polynomial chaotic potential.  For each curve, $A_s$ is fixed while
the branch-dependent mass $m(n_s)$ is re-determined self-consistently.
The larger-$m$ branch lies near $N_T\simeq129.7$, while the smaller-$m$
branch gives $N_T\simeq131.2$--$131.4$ at the central observational
points.  The narrow feature near $n_s\simeq0.96967$ on the smaller-$m$
branch marks a separatrix between trajectories with opposite signs of
$\dot\phi_B$.}
\label{fig:poly_NT}
\end{figure}

For the total-expansion curves, the construction is different from the
reheating scan just described.  In Fig.~\ref{fig:poly_NT}, $A_s$ is
held fixed for each data set, but the slow-roll equations are solved
again at every $n_s$, so that the appropriate branch-dependent mass
$m(n_s)$ is re-determined self-consistently before the trajectory is
evolved backward to the mLQC-I bounce.

\begin{table*}[!ht]
\centering
\caption{Central total-expansion values for the two branches of the
polynomial chaotic potential and the corresponding LQC comparison.}
\label{tab:poly_NT}
\begin{tabular}{llccc}
\hline\hline
Data set
& branch
& $N_T^{\rm mLQC-I}$
& $N_T^{\rm LQC}$
& $\Delta N_T$
\\
\hline
Planck 2018
& $\phi_*>0$
& 129.72833
& 129.96852
& $-0.24019$
\\
Planck 2018
& $\phi_*<0$
& 131.20024
& 131.44044
& $-0.24019$
\\
ACT 2025
& $\phi_*>0$
& 129.73163
& 129.97183
& $-0.24019$
\\
ACT 2025
& $\phi_*<0$
& 131.35113
& 131.59132
& $-0.24019$
\\
Combined
& $\phi_*>0$
& 129.73233
& 129.97252
& $-0.24019$
\\
Combined
& $\phi_*<0$
& 131.20449
& 131.44469
& $-0.24019$
\\
\hline\hline
\end{tabular}
\end{table*}

Table~\ref{tab:poly_NT} and Fig.~\ref{fig:poly_NT} show that the
two observational branches shift the absolute value of the total
expansion by about $1.5$ e-folds.  Nevertheless, both remain close to
the characteristic scale $N_T\sim130$.  Even more strikingly, the
difference between LQC and mLQC-I is essentially the same on both
branches and for all three observational data sets:
\begin{equation}
\Delta N_T
=
N_T^{\rm mLQC-I}-N_T^{\rm LQC}
\simeq
-0.24019.
\label{eq:DeltaNT_poly}
\end{equation}

A higher-resolution scan of the smaller-$m$ branch reveals that the
narrow feature in Fig.~\ref{fig:poly_NT} is a genuine separatrix rather
than a numerical artifact.  For the Planck normalization its location is
\begin{equation}
 n_s^{\rm sep}\simeq0.96967184,
\end{equation}
with shifts below $6\times10^{-11}$ when the Planck, ACT, and combined
values of $A_s$ are used.  Across the separatrix the sign of the bounce
velocity $\dot\phi_B$ changes.  As $n_s$ approaches
$n_s^{\rm sep}$ from either side, the bounce-to-pivot expansion grows
rapidly; over the numerically resolved range it is approximately linear
in $-\log_{10}|n_s-n_s^{\rm sep}|$.  The location of the feature is
stable when the switch from the post-bounce Friedmann equation to the
canonical mLQC-I system is varied over
$0.1\leq\rho/\rho_c^{\rm I}\leq0.99$ and when the integration
tolerance is tightened from $10^{-9}$ to $10^{-11}$.  This behavior is
therefore associated with the phase-space structure of the smaller-$m$
branch and is highly localized in $n_s$; it does not affect the central
observational values quoted in Table~\ref{tab:poly_NT}.

Taken together, the PLP, Starobinsky, and polynomial chaotic examples
show two distinct levels of model dependence.  The reheating quantities
$N_{\rm re}$ and $T_{\rm re}$, as well as the location of instantaneous
reheating, can depend strongly on the shape of the potential and, in the
polynomial model, on the choice of observational branch.  The absolute
value of $N_T$ also displays some potential and branch dependence, but
the characteristic bounce-to-present expansion remains of order
$N_T\sim130$.  Most notably, within the leading-order treatment
${\cal{F}}_{\rm I}(k_*)=1$, all three potentials considered here yield nearly
the same LQC--mLQC-I shift,
\begin{equation}
\Delta N_T\simeq-0.24.
\label{eq:DeltaNT_summary}
\end{equation}
This remarkable agreement suggests that the shift is governed primarily
by the difference between the LQC and mLQC-I background geometries,
rather than by the detailed form of the inflationary potential.  A
broader survey of viable inflationary models would be needed to
establish how general this behavior is.

\FloatBarrier
\section{Conclusions and Outlook}
\label{sec:conclusions}

In this paper we have studied the reheating epoch and the total
bounce-to-present expansion in modified loop quantum cosmology I
(mLQC-I).  Our analysis was carried out using the effective post-bounce
dynamics of mLQC-I and current observational determinations of
$(A_s,n_s)$ from Planck 2018, ACT 2025, and the combined
Planck+ACT+DESI DR2 data set.  Throughout the analysis we adopted the
leading-order approximation ${\cal{F}}_{\rm I}(k_*)=1$ for the primordial
spectrum at the observational pivot scale.

We first considered the power-law plateau (PLP) potential.  For each
observational data set, the potential normalization was determined
directly from $(A_s,n_s)$ rather than treated as an independent free
parameter.  Although the resulting values of $V_0$ differ appreciably,
the quantum-bounce regime is remarkably insensitive to this variation.
For kinetic-energy-dominated initial conditions, the post-bounce
solutions associated with the three observational normalizations are
practically indistinguishable near the bounce and begin to separate only
later, when the potential becomes dynamically important.  The
inflationary duration is therefore more sensitive to the observational
normalization than the genuinely quantum part of the evolution.

Using the standard effective description of reheating in terms of a
constant equation-of-state parameter $w_{\rm re}$, we then calculated
the reheating e-fold number $N_{\rm re}$ and the reheating temperature
$T_{\rm re}$.  For the PLP potential, the instantaneous-reheating point
occurs near
\[
n_s^{\rm inst}\simeq0.970.
\]
The Planck and ACT central values lead to similar reheating histories
for the representative cases considered here, whereas the combined
central value lies on the opposite side of the instantaneous-reheating
point.  In particular, among the four values
$w_{\rm re}=-1/3,0,2/3,$ and $1$, the combined data allow the
$w_{\rm re}=1$ case to remain above the BBN temperature bound, while
the $w_{\rm re}=2/3$ case falls below it.

A central result of the paper is the total number of e-folds from the
mLQC-I bounce to the present.  After combining the reheating relations
with entropy conservation, the explicit dependence on
$N_{\rm re}$ and $w_{\rm re}$ cancels, leaving the simple relation
\[
N_T
=
\ln\left(\frac{a_*}{a_B}\right)
+
\ln\left(\frac{H_*}{k_*^{\rm ob}}\right).
\]
For the PLP potential, the three central observational data sets give
$N_T\simeq129.43$--$129.46$, and throughout the interval
$0.955\leq n_s\leq0.985$ we find
$N_T\gtrsim129.41$, or conservatively $N_T\gtrsim129$, with a
characteristic total-expansion scale $N_T\sim130$.
A direct comparison with LQC under the same assumptions shows a small
but systematic shift,
\[
\Delta N_T
\equiv
N_T^{\rm mLQC-I}-N_T^{\rm LQC}
\simeq -0.24.
\]

To test the sensitivity of these conclusions to the inflationary
potential, we repeated the analysis for the Starobinsky potential and
for a polynomial chaotic potential.  The reheating predictions were
found to be strongly model-dependent.  For the Starobinsky potential,
the instantaneous-reheating point shifts to
$$
n_s^{\rm inst}\simeq0.9653. 
$$
This change is already sufficient to
alter which of the representative values of $w_{\rm re}$ give positive
$N_{\rm re}$ at the ACT central value.  Moreover, at the combined
central value none of the four representative constant-$w_{\rm re}$
cases considered here remains compatible with the BBN temperature
bound.

The polynomial chaotic potential exhibits an additional two-branch
structure.  For a fixed observational pair $(A_s,n_s)$, the slow-roll
equations admit two relevant solutions for $\phi_*$ and therefore two
different values of the mass parameter $m$.  The two branches lead to
substantially different values of $N_*$ and to qualitatively different
reheating histories.  For example, at the Planck central value the
larger-$m$ branch admits viable reheating for
$w_{\rm re}=-1/3$ and $w_{\rm re}=0$, whereas on the smaller-$m$
branch these cases have negative $N_{\rm re}$ and only the
$w_{\rm re}=1$ case among the four representative choices remains
above the BBN bound.  This demonstrates that reheating can be sensitive
not only to the functional form of the potential but also to the
observational branch realized by the same potential.
In addition, a higher-resolution analysis of the smaller-\(m\) branch also reveals a highly localized separatrix near \(n_s^{\rm sep}\simeq0.96967184\), across which the sign of \(\dot{\phi}_B\) changes; this feature does not affect any of the central observational values considered here.

In contrast, the total expansion is considerably more robust.  The
Starobinsky potential gives $N_T\simeq129.06$ at the three central
observational points, while the two polynomial branches give
$N_T\simeq129.7$ and $N_T\simeq131.2$--$131.4$, respectively.  Thus the
absolute value of $N_T$ does display some potential and branch
dependence, but in all cases studied here the bounce-to-present
expansion remains at the characteristic scale
$N_T\sim130$.
More remarkably, the LQC--mLQC-I difference remains nearly unchanged:
\[
\Delta N_T\simeq-0.24,
\]
for the PLP, Starobinsky, and both branches of the polynomial chaotic
potential.  The persistence of this value across the three potentials
considered here suggests that the shift is controlled primarily by the
difference between the LQC and mLQC-I background geometries rather than
by the detailed form of the inflaton potential.  A broader survey of
inflationary models would be required before assigning this behavior a
more general status.

Several qualifications should be kept in mind.  First, the present
analysis uses the leading-order approximation
${\cal{F}}_{\rm I}(k_*)=1$ for the primordial spectrum.  An explicit calculation
of ${\cal{F}}_{\rm I}(k)$ 
\cite{Li:2021mop,Pan:2025ppa} would be valuable for
assessing the size of the neglected perturbative corrections.  Second, reheating has been modeled by a
constant effective equation-of-state parameter and by a small set of
representative values of $w_{\rm re}$; no specific microscopic
reheating mechanism has been assumed.  Consequently, the reheating
constraints obtained here should be interpreted as effective
phenomenological bounds rather than as a determination of a unique
reheating history.

Finally, the present work has focused on mLQC-I.  It would be
interesting to extend the potential-dependence analysis to the other
loop-quantized cosmological models.  In particular, a systematic study
of the sensitivity of $N_{\rm re}$, $T_{\rm re}$, and $N_T$ to the
inflationary potential in LQC and mLQC-II would provide a useful test of
which features are controlled by the choice of potential and which are
genuine signatures of the underlying quantization prescription.  We
leave this comparison to future work.

\begin{acknowledgments}

A.W. thanks Drs. Bao-Fei Li, Mayukh R. Gangopadhyay, and Yogesh for their earlier collaboration on reheating within the framework of LQC. J.S. and R.P. thank Baylor University for generous support through the Physics $\&$ Astronomy graduate program.  A.W. is partially supported by the U.S. National Science Foundation under Grant No. PHY-2308845. We also acknowledge the use of ChatGPT as a computational
aid in performing some mathematical calculations.

\end{acknowledgments}


\bibliographystyle{apsrev4-1}
\bibliography{References_LQCs.bib}

\end{document}